\documentstyle[preprint,aps]{revtex}
\tightenlines
\begin{document}

\draft

\title{Collective charge density excitations in two-component 
one-dimensional quantum plasmas: Phase fluctuation mode dispersion 
and spectral weight in semiconductor quantum wire nanostructures}
\author{S. \ Das Sarma and E. H.\ Hwang}
\address{Department of Physics, University of Maryland, College Park, 
Maryland   20742-4111} 
\date{\today}
\maketitle

\begin{abstract}

Collective  charge density excitation spectra of both  a
spatially separated two-component 
quasi-one dimensional (1D) quantum plasma, as existing in a
semiconductor double quantum wire structure, and a 1D homogeneous 
electron-hole plasma, as appropriate for a photoexcited semiconductor 
quantum wire system,  are 
calculated within the two-component random-phase-approximation.
We find two phase fluctuation collective modes, one (optical plasmon, OP) 
with energy proportional to $q|\ln(qa)|^{1/2}$ is the {\it total} in-phase 
(out-of-phase) charge density oscillation of the system  
and the other (acoustic plasmon, AP) with a linear energy dispersion as
$q \rightarrow 0$ is the {\it neutral} out-of-phase (in-phase) charge density 
oscillation of the system for the situation where the two components have 
the same (opposite) charges, where $q$ is the 1D wave vector and $a$ 
is a characteristic 1D confinement size.  
In contrast to higher dimensional systems we find the neutral long 
wavelength AP mode to be generically undamped by Landau damping effects 
due to the severe suppression of single particle excitations in 1D systems.
We also investigate the effect of impurity scattering on the
collective mode dispersion and damping, and calculate the collective 
mode spectral weight by obtaining the dynamical structure factor. 
We find that both OP and AP modes are
overdamped by impurity scattering below some critical wave vector.
We find that in the long wavelength limit the spectral
weight is carried mostly by the OP, but  the spectral
weight of the undamped AP mode at finite (but not too large) wave
vectors is comparable 
with that of the OP mode, making it viable to observe the
AP mode in semiconductor quantum wire systems.
The effect of the interwire electron tunneling in a biwire system on
the collective charge density excitation spectra is also studied.
We discuss the mode dispersion and damping from an effective Luttinger 
liquid perspective as well and in some cases include local field 
corrections in our calculations.


\end{abstract}

\newpage

\section{introduction}

Recent advances in fabrication techniques involving molecular beam
epitaxy and lithography have now made it possible to make narrow
GaAs-based quasi-one dimensional (1D) electronic systems
with lateral dimensions of the
order of the Bohr radius \cite{reed}.
In these so-called quantum wire structures, the motion of charge
carriers is confined in two transverse directions but
is essentially free (in the effective mass sense) in the longitudinal
direction.  Study of 
collective modes in reduced dimensional electron  systems in 
semiconductor nanostructures is a subject of growing experimental and
theoretical interest.
Experimentally, far-infrared  optical
spectroscopy\cite{deme,hans1} and resonant
inelastic light scattering spectroscopy\cite{pinc,egel,goni} have been
used to study quasi-1D
elementary electronic excitations.
Several theoretical studies, mostly based on the random phase
approximation (RPA), have been reported on the energy
dispersion of elementary excitations in semiconductor quantum wires.
The measured 1D intrasubband plasmon 
dispersion agrees remarkably well with the
RPA predictions \cite {li}. The quantitative agreement
between RPA predictions and the  experimental
results \cite{goni} was later explained 
by the fact \cite{li2} that the calculated RPA plasmon dispersion and the 
Tomonaga-Lutinger  theory  for  the  collective charge density
excitations  of the 1D electron   
system are equivalent at long wavelengths by virtue of the vanishing of vertex 
corrections \cite{dl} in the irreducible polarizability of the 1D
electron gas. 

The two-component quasi-1D systems which are composed of electrons and
holes have been generated in a wide variety
of semiconductor quantum wires by optical pumping. 
Spatially separated two-component quasi-1D systems can be produced by
additional lateral confinement to produce a double quantum wire
structure in 
GaAs--Ga$_x$Al$_{1-x}$As double quantum well systems.
The collective modes of the two-component plasma also
play important roles in the many-body physics of carriers
such as screening of the Coulomb interaction potential and in Coulomb
drag problem.
In this paper, we calculate the dispersion and the spectral weight of
longitudinal collective modes (with and without spatial separation)
in quasi-1D two-component
plasmas as in semiconductor double quantum wire structures or
in photoexcited homogeneous 1D electron-hole plasmas (EHP). Our theory is  
based mostly on the two-component RPA (which should be well-valid in 1D 
systems); in the 1D EHP case we go beyond RPA to include local field 
corrections to assess the validity of 1D RPA.
The collective modes of the two-component
plasma have been widely studied in higher dimension \cite{pines,pinczuk,dsm},
and recently, numerically in quasi-1D EHP \cite{tan}.

It is well-known that a two-component plasma has two branches of
longitudinal collective excitation spectra called the optical plasmon (OP)
and the acoustic
plasmon (AP),  where the density fluctuations in each component oscillate
in-phase (OP) and out-of-phase (AP) respectively relative to each other,
assuming the two charge components to be the same.
As expected, we find two plasma modes in the two-component 1D system;
the in-phase OP with a
typical 1D plasmon dispersion with the frequency 
proportional to
$q|\ln(qa)|^{1/2}$ at long 
wavelengths and the out-of-phase AP with its frequency
linear in $q$ as $q \rightarrow 0$, where $q$ is the 1D wave vector and 
$a$ is a characteristic 1D confinement size (i.e., roughly the wire width).
Unlike 2D and 3D systems, where the electron-hole pair continua have
only an upper boundary, 
$qv_F +q^2/2m$, the 1D pair continuum has also a low energy gap
under the lower boundary $|qv_F -q^2/2m|$, implying no low energy single
particle excitations (SPE) are allowed in 1D.  This gap,  the
non-existence of low energy SPE in 1D Fermi
systems, arises from
severe phase space restrictions (Pauli exclusion principle)
imposed  by 1D energy-momentum conservation, and leads 
to the possibility of the existence of an undamped plasmon
mode in multi-component 1D systems\cite{hwang}.
We find a low energy undamped AP mode in the
double-quantum-wire system for arbitrary interwire separations 
(including the zero separation case).
In general, in higher dimensions, the collective mode spectral weight
is mostly carried 
by the OP, which makes the experimental observation of the AP mode in
higher dimensions rather difficult. In contrast, we find that in quasi-1D
systems  the spectral weight of the undamped AP mode is comparable
to that of the OP mode and the spectral weight of the AP mode
increases as the spatial 
separation increases, which is similar to the corresponding 2D situation.
The greater relative spectral weight of the AP
mode in 1D is directly related to the non-existence of low energy 1D
SPE.

The plasma modes in 1D systems have vanishing frequencies at long
wavelengths, and in general their energies at experimentally
accessible wave vectors are small ($\approx 1$ meV). Their
observability, therefore, depends rather crucially on the absence of
appreciable damping or broadening in the system. While the absence of
long wavelength Landau damping (to SPE) makes it feasible to discuss
the observability of the AP mode, one must also have very small
impurity broadening in order for the plasmon modes to be
observable. Impurity scattering causes the
carrier motion to be diffusive assuming the scattering to be weak.
In principle, any impurity disorder, no matter how weak, localizes all 
single electron 1D states; for weak disorder, however, the localization 
length is larger than the system size and electron dynamics could be 
considered to be diffusive.
The diffusive nature of the
electronic dynamics strongly affects the plasmon
dispersion in the long wavelength limit because the plasma frequency is 
vanishingly small at long wavelengths. The 1D plasmons therefore become
overdamped due to impurity scattering induced level broadening at sufficiently long wavelengths. This
plasmon level broadening may actually 
play a significant role in determining the physical characteristics 
in quantum wires. 
The collisional damping of the plasmon modes due to
impurity scattering occurs in addition to the Landau damping which
arises from the decay of the plasmon to SPE electron-hole pair excitations. 
(As emphasized above, while Landau damping is suppressed in 1D,
impurity broadening is always present unless the system is perfectly
pure.) In the single-component plasma, it is known 
that, as the impurity scattering increases, the collisional damping becomes
strong and the plasmon mode becomes overdamped and disappears at small
$q$ in 1D and 2D systems (by contrast, this is a small effect in 3D
since the plasmon energy is finite at zero wave vector) \cite{dashwang}.
Inclusion of impurity scattering in 1D and 2D overdamps the plasmon below
a critical wave vector \cite{dashwang,hu}. 
This disappearance of the plasmon spectral weight at small $q$ has
important consequences for the existence and characteristics of the
Fermi surface in
a one dimensional electron gas \cite{hu}. 
In this paper we  investigate  the
plasmon damping of the two-component 1D plasma in the presence of
impurity scattering. (As noted above any finite
impurity disorder leads to Anderson localization of 1D one electron
states, and therefore the 1D electron dynamics in the presence of
impurity disorder is exponentially localized, {\it not} diffusive --
this is, however, a purely academic point of little relevance to high
quality GaAs quantum wire systems with low disorder, where the
localization lengths are many microns long and are in fact longer than
the 1D sample lengths \cite{liu}; we are therefore justified in
treating the effective electron dynamics as being diffusive,
neglecting the Anderson localization effects in high quality quantum
wires of relevance to plasmon experiments.)  

We also
discuss the effect of interwire electron tunneling on the bi-wire
collective mode spectra.
The most important qualitative feature of the plasmon dispersion in
the presence of interwire tunneling 
is that, even though the in-phase OP mode $\omega_+$  depends weakly on the
tunneling, the out-of-phase AP mode $\omega_-$, which
is purely acoustic in the absence of tunneling ($\omega_- \sim q$),
develops a plasmon gap at $q=0$ in the presence 
of nonzero tunneling; the plasmon gap in the AP
depends nontrivially on the 1D electron density $n$, the interwire
tunneling amplitude $t$, and the interwire distance $d$. 
Tunneling, therefore, effectively changes the AP ($\omega_- \sim q$)
into an optical mode with a long wavelength gap.

We use  the RPA for most of our calculations. As emphasized, RPA is a
very good approximation for the collective mode dispersion
in 1D quantum wire structures by virtue of the essential vanishing
of all vertex corrections to the 1D polarizability function.
We consider the zero temperature case and assume that our quasi-1D
system has an  infinite
square-well confinement with a finite width $a$ in the 
y-direction  and a zero width in the z-direction. It is easy to
include a finite width in the $z$ direction but our results do not
change qualitatively. 
In typical quantum wires, the wire width along the
growth direction (z-direction) is much smaller (by an order of
magnitude) than that in the
lateral direction (y-direction), justifying our 2D finite width model
for the 1D wire. 
In the finite-width 2D model we use a square well confining potential
with infinite 
barriers at $y=-a/2$ and $y=a/2$, with the electron dynamics along the
$x$-axis being free. 
Confinement of the electrons in the $y$ and $z$ 
directions leads to the quantization of energy levels into different
1D subbands, and we assume the 1D extreme quantum limit where only the
lowest subband is occupied by carriers, and all the 1D higher subbands
are neglected.
We study the plasmon modes of the spatially separated two-component
system by taking two identical parallel quantum wires separated
by a distance $d$.
We include results for the case $d=0$ because it applies directly to
the photoexcited 1D EHP where electrons and holes are not spatially separated.
We consider two different spatial configurations for our two-component
bi-wire structure: two 1D wires  with 
a spatial separation $d$ being parallel to each other aligned in the
(1) $x-z$ plane (separated along the $y$-axis) and (2) $x-y$ plane
(separated along the $z$-axis). Since the difference in our calculated 
results between these two configurations is within one percent, 
we only give the results for the configuration (1).

It is well-known that 1D electron systems are fundamentally different 
from their higher dimensional counterparts because 1D interacting 
electrons form a Luttinger liquid, not a Fermi liquid, and any
electron-electron interaction, no matter how weak, destroys the 
Fermi surface, i.e., the discontinuity in the momentum distribution 
function, in 1D. It turns out, however, that this distinction, while 
being profound at a fundamental theoretical level (because it implies 
the non-existence of quasiparticles in 1D systems), is rather irrelevant
to the understanding of the collective mode spectra and their 
experimental realization in semiconductor quantum wire structures. 
As has been emphasized elsewhere, RPA gives a very good account of 
the collective mode properties in 1D semiconductor quantum wire 
structures, and we discuss this issue in some details in this paper 
in the context of the two-component 1D plasma.

The rest of the paper is organized as follows: In Sec. II we provide 
the basic RPA formalism as well as  the calculated analytic and 
numerical results
for plasmon  dispersion, damping, and spectral weight
by obtaining the dynamical structure factor 
(which is a direct measure of the light scattering 
intensity); in Sec. III we study the effects of the interwire electron
tunneling in a bi-wire system on 
the collective charge density excitation spectra; in Sec. IV the
plasmon  dispersion and  damping in the photoexcited 1D EHP is
studied in both RPA and including local field corrections; 
in Sec. V we briefly discuss the acoustic plasmon mode 
of the two-component 1D Luttinger liquid;
we conclude in Sec. VI providing a brief summary of our
results.

\section{RPA Theory and results}

In general, the density fluctuation spectra or the plasmons or the
longitudinal collective modes of  a solid state 
plasma are given by the poles of the
density-density correlation function, or equivalently by the zeros of
the dynamical dielectric function. 
For a multicomponent plasma the collective modes 
are given by the zeros of the generalized dielectric tensor
$\epsilon$. In the two-component spatially separated 1D plasma the
generalized dielectric tensor 
with wire indices {\it i, j} ($=1,2$ indicating the two quantum wires)
is given (neglecting any inter-wire tunneling) within the RPA by
\begin{equation}
\epsilon_{ij}(q,\omega) = \delta_{ij}-V_{ij}(q)\Pi_{jj}
(q,\omega),
\label{epsi}
\end{equation}
where $q$ is the 1D wave vector in the $x$-direction of free motion,
$\omega$ is the mode frequency ($\hbar =1$ throughout this paper),
$V_{ij}(q)$  the Coulomb  
interaction in the 1D wire representation, and $\Pi_{jj}(q,\omega)$  the 
1D non-interacting irreducible
polarizability function for the $j$-th ``component'' (i.e., the $j$-th
wire with $j=1,2$ corresponding
to the two wires).
In an ideal 1D electron system the
Coulomb interaction  $V(q)$ in the wave vector space is
logarithmically divergent, but in our  realistic
finite width quantum wire model with the confinement width $a$ we obtain the
intra-wire [inter-wire] Coulomb interaction matrix element
$V(q)$ [$U(q)$] by taking the quantizing
confinement potential to be of infinite square well type\cite{hu},
\begin{equation}
V(q) = V_{11}(q)=V_{22}(q) = \frac{2e^2}{\epsilon_b} f(q),
\label{vq}
\end{equation}
\begin{equation}
U(q) = V_{12}(q)=V_{21}(q)=\pm \frac{2e^2}{\epsilon_b} g(q),
\label{uq}
\end{equation}
where $\epsilon_b$ is the background high-frequency lattice dielectric
constant, 
the $+$ ($-$) sign in Eq. (\ref{uq}) 
refers to interaction between the same
(opposite) types of charge in the two wires, (the two wires 
could in principle have electrons and
holes respectively, rather than being just all electrons or 
all holes  in both wires), and
$f(q)$, $g(q)$ are form factors associated with confinement matrix 
elements, which are given, for our infinite square well confinement model, by
\begin{equation}
f(q) =\int^{1}_{0}dx K_0(|qax|) h(x),
\end{equation}
\begin{equation}
g(q) = \int^{1}_{0}dx K_0 \left (q \sqrt{(ax)^2 + d^2} \right ) h(x),
\end{equation}
for two wires parallel to each other aligned in the $x-z$ plane. 
Here $K_0$ is the zeroth-order modified Bessel function of the second
kind, and the function $h(x) = (1-x) \left [ 2+\cos(2\pi x) \right ] +
3\sin(2\pi x)/(2\pi)$. 
If the two wires are parallel to each other aligned in the $x-y$ plane the
inter-wire Coulomb form factor becomes [with $f(q)$  remaining the same]
\begin{equation}
g(q) = \int^{1}_{0}dx K_0 \left (|q (ax + d)| \right ) h(x).
\end{equation}
As $q\rightarrow 0$, $\;\;f(q),g(q) \sim K_0(|qa|) + C$, where $C$
is a constant (independent of $d$) for $f(q)$, but a function of the
inter-wire spatial separation $d$ for $g(q)$. 
Note that as $d \rightarrow 0$ $U(q)$ becomes $V(q)$, as it should,
because the two wires are identical.

The irreducible 1D polarizability $\Pi_0(q,\omega)$ in the pure 1D system
is  given by the bare bubble diagram. The analytic form of
$\Pi^0_{ii}(q,\omega)$ for complex frequency is given by\cite{li}
\begin{equation}
\Pi^0_{ii}(q,\omega)=\frac{m_i}{\pi q} \ln \left [ \frac{ \omega^2 -
(q^2/2m_i - q 
v_{Fi})^2} {\omega^2 - (q^2/2m_i + qv_{Fi})^2} \right ],
\label{pi0}
\end{equation}
where the principal value of logarithm (i.e., $ -\pi < 
\mathrm{Im} [\ln(z)]
< \pi$) should be taken for complex frequency. In evaluating
$\Pi_0(q,\omega)$ for real 
frequency, the usual retarded limit $\omega \rightarrow \omega + i0^+$ 
is implied.
In Eq. (\ref{pi0}) $m_i$ and $v_{Fi}$ are the effective band mass and
the Fermi velocity, respectively. 
In the long wavelength limit ($q \rightarrow 0$), we can expand 
the irreducible 1D polarizability as
\begin{equation}
\Pi^{0}_{ii}(q,\omega) \approx \frac{2 v_{Fi}}{\pi} \frac{q^2}{w^2 -
(q v_{Fi})^2}, 
\label{fix}
\end{equation}
where  $v_{Fi}=k_{Fi}/m_i$ is the Fermi velocity with a Fermi momentum
$k_{Fi}=\pi n_i/2$ for the  $i$-th 1D component with $n_i$ being the 
1D density in the $i$-th component.
Since RPA is known to be valid in the long-wavelength limit ($q\rightarrow
0$), the limiting forms for the polarizability would be sufficient for
our analytic calculations 
since we are interested in the leading order wave vector dependence of
the collective modes of the system. [For our numerical results we use
the full 1D $\Pi(q,\omega)$ as given in Eq. (\ref{pi0}).]
Note that $\Pi$, $v_F$, $k_F$, etc., refer to the noninteracting or 
the bare 1D system, and are therefore perfectly well-defined.

The expression for the 1D polarizability $\Pi_0(q,\omega)$ given in
Eq. (\ref{pi0}) assumes that the system is pure and free
from any impurity scattering.
In the presence of impurity scattering characterized by a level
broadening or damping $\gamma$, the polarizability
$\Pi_{\gamma}$ can be
evaluated by means of the standard perturbation theory \cite{mermin}
to include the impurity scattering induced vertex correction. 
The impurity scattering 
effects are introduced diagrammatically in
the RPA by including impurity ladder diagrams
into the electron-hole bubble consistently with self-energy
corrections in the electron Green's function. 
Since the exact expression for $\Pi_{\gamma}(q,\omega)$ within this
diagrammatic approach is complicated,
we use in the numerical calculation a particle-conserving
approximation for arbitrary values of $q$ and $\omega$,
given  by Mermin \cite{mermin}. The Mermin expression is, in fact,
equivalent to the diagrammatic result at long wavelength as long as
the density of states renormalization by impurity scattering is
negligible. In this relaxation time
approximation\cite{mermin}, with the impurity scattering induced level
broadening $\gamma=1/(2\tau)$, the polarizability is given by
\begin{equation}
\Pi_{\gamma}(q,\omega) = \frac{ (\omega + i \gamma) \Pi_0(q, \omega +
i \gamma)} { \omega + i \gamma [ \Pi_0(q, \omega+i\gamma)/ \Pi_0(q,0)]}.
\label{pig}
\end{equation}
For small $q$ and $\omega$, the polarizability within the Mermin
formula becomes
\begin{equation}
\Pi_{\gamma}(q,\omega)\approx \frac{nq^2}{m\omega(\omega+i\gamma)}.
\label{piga}
\end{equation}
This approximate formula gives the same  long wavelength and
low frequency diffusive behavior as one gets from the
diagrammatic approach \cite{dashwang}. 
In this paper, we take the impurity scattering
induced broadening $\gamma$ as a constant phenomenological parameter,
which, for example, could be taken from the experimentally measured
carrier mobility $\mu = e^2 \tau/m$, with $\gamma = 1/(2\tau)$.

The condition for the existence of a collective mode is given by the
zeros of $|\epsilon|$, the determinant of the dielectric tensor
defined by Eq. (\ref{epsi}).
For a two-component system without any tunneling, we have within RPA:
\begin{equation}
1-V(q) \left [ \Pi^0_{11}(q,\omega) + \Pi^0_{22}(q,\omega) \right ] +
\left [ V(q)^2 - U(q)^2 \right ]
\Pi^0_{11}(q,\omega)\Pi^0_{22}(q,\omega) =0,
\label{zeros}
\end{equation}
with $i,j=1,2$ being the two charge components in confined wires 1 and 2.
This equation indicates that the sign of the charge components forming
the plasma has no effect on the dispersion relation of the collective
modes of a two-component system. 

The analytical formulae for the long wavelength
plasmon dispersion can be obtained even in the general situation where
the two charge components have different Fermi wave vectors and Fermi
energies (by virtue of having, for example, different effective masses
$m_{1,2}$ and densities $n_{1,2}$).
In this section we study  the collective modes in this general
situation (i.e., $d \neq 0$, $n_1 \neq
n_2$, and $m_1 \neq m_2$) although we should emphasize that such
a completely general situation is not physically particularly
applicable (because in most existing quantum wire systems either one has 
$m_1=m_2$ and $d \neq 0$ or $m_1 \neq m_2$ and $d =0$, corresponding 
respectively to a bi-wire system or an EHP system), and in Sec. IV, 
we investigate, in detail, the modes in
a photoexcited homogeneous (i.e., $d=0$) 1D EHP ($d=0$, $n_1=n_2$, 
$m_1 \neq m_2$)  where
the electrons and holes are not spatially separated, but have different 
effective masses.

First, we consider the plasmons of a symmetric two-component system having 
equal densities ($n_1 = n_2 = n$), masses ($m_1 = m_2 = m$), and a
finite inter-wire separation $d$, and
neglecting impurity scattering 
effects ($\gamma = 0$) for simplicity. In this case 
$\Pi_{11}^0 = \Pi_{22}^0$ and
Eq. (\ref{zeros}) becomes
\begin{equation}
\left[1- 2 V_+(q) \Pi_{11}^0 (q,\omega) \right ] \times
\left[1- 2 V_-(q) \Pi_{11}^0 (q,\omega) \right ] =0,
\label{zeros1}
\end{equation}
where $V_{\pm} = [V(q) \pm U(q)]/2$.
By solving Eq. (\ref{zeros1}) we have two branches of 
collective modes corresponding to the in-phase OP mode
$\omega_+$ and the out-of-phase
AP mode $\omega_-$ (the in-phase $\omega_+$ and
out-of-phase $\omega_-$ nomenclature applies to the situation when
both components are electrons or both are holes; otherwise, i.e., 
for an electron-hole system, $\omega_+$
is the out-of-phase mode and $\omega_-$ the in-phase mode):
\begin{equation}
\omega_{\pm}^2(q) = \frac{A_{\pm}(q) \left [ v_F q + E(q)
\right ]^2 - \left [ v_F q - E(q) \right ]^2}{A_{\pm}(q)
-1},
\label{wpm}
\end{equation} 
where $E(q) = q^2/2m$ and $A_{\pm}(q) = \exp[\pi q/(2mV_{\pm})]$. 
The optical mode $\omega_+$ (acoustic mode $\omega_-$) indicates the
symmetric in-phase (antisymmetric out-of-phase) density oscillation of the
coupled system assuming the two carriers have the same charge. In
the long wavelength limit $q \rightarrow 0$,  Eq. (\ref{wpm}) becomes
\begin{equation}
\omega_{\pm}(q\rightarrow 0) = q \left [ v_F^2 + \frac{4}{\pi}v_F
V_{\pm}(q \rightarrow 0) \right ]^{1/2},
\end{equation}
where the asymptotic forms of the Coulomb potential $V_{\pm}(q)$ as $q
\rightarrow 0$ are given by
\begin{eqnarray}
V_+(q \rightarrow 0) & = & \frac{2e^2}{\epsilon_b} \left [K_0(|qa|) +
1.97269... \right ], \nonumber \\
V_-(q \rightarrow 0) & = & \frac{2e^2} {\epsilon_b}\int_0^1dx \ln \left [
\frac{y(x)}{x} \right ]\left [(1-x)\,(2+\cos(2\pi x)) +
\frac{3}{2\pi}\sin(2\pi x) \right ],
\end{eqnarray}
where $y(x) = [x^2 + (d/a)^2]^{1/2}$ \ \ $\left [ y(x) = x + d/a \right ]$
for the system 
aligned in the {\it x-z} ({\it x-y}) plane. 
As $q\rightarrow 0$, $V_+(q)$diverge as $-\ln(qa)$, but
the leading order term in $V_-(q)$ is finite and depends only on
the spatial separation $d$. Therefore, the leading order term in the
1D OP dispersion $\omega_+(q)$ is enhanced  by a factor
$\sqrt2$ compared to the one-component system (corresponding 
trivially to the change in density from $n$ to $2n$) and behaves as
$q |\ln(qa)|]^{1/2}$ as $q\rightarrow 0$.
The dispersion of the acoustic mode $\omega_-(q)$ shows a purely 
linear behavior $\omega_-(q) = v_-q$ for $q \rightarrow 0$ with the 
velocity $v_- = v_F[1 + 4 V_{-}(q =0)/(\pi
v_F)]^{1/2} $ as $q\rightarrow 0$, and lies always above the single particle
Landau  damping region because $v_- > v_F$.
The spatial separation of the two wires produces a weak 
second order correction to the OP dispersion in the
two-component system.
For $d=0$, the $\omega_-$ mode becomes degenerate
with the electron-hole continuum (i.e., $v_- = v_F$) in the long 
wavelength limit,
and the $\omega_+$ mode becomes that of the corresponding 
one-component system with a total density $N=2n$. 
As the spatial separation increases the optical (acoustic) plasmon mode
frequency decreases (increases) slowly at long wavelengths.  
For the system with an infinite separation, the two modes become 
uncoupled, and have the
same dispersion as the 1D plasmon in the
one-component system with a total density $n$.

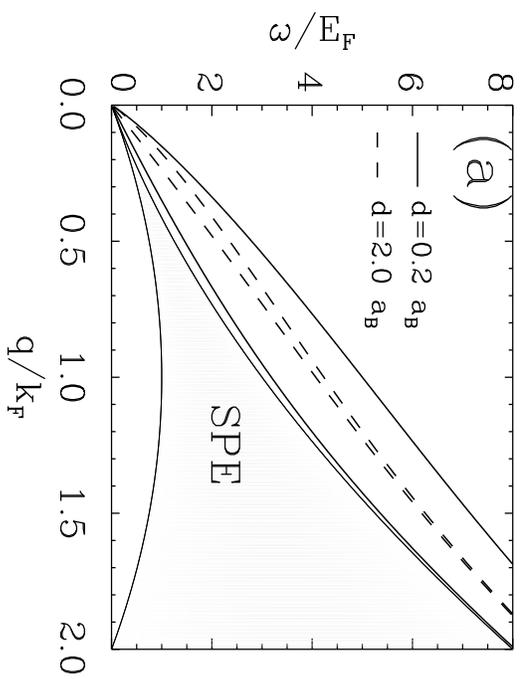
\begin{figure}
\caption{ 
Calculated RPA collective mode dispersion for 
different spatial separations between the two identical quantum wires;
(a) $n_1 = n_2 = 0.6\times 10^6 \; {\rm cm}^{-1}$ and
(b) $n_1 = n_2 = 10^6 \; {\rm cm}^{-1}$.
Here, the higher (lower) frequency  mode corresponds to the OP
(AP) mode in each case. 
Shaded region indicates the single particle excitation (SPE) continuum.
}
\label{fig0}
\end{figure}

In Fig. \ref{fig0} we show our calculated collective modes dispersion for
different spatial separations between the two identical quantum wires; 
(a) $d=0.2 a_B$ and $2.0 a_B$ with $n = 0.6\times 10^6 \; {\rm cm}^{-1}$, 
and (b) $d=2.0 a_B$ and $4.0 a_B$ with $n = 10^6 \; {\rm cm}^{-1}$,
where $a_B$ ($\approx 100 \; \AA$) is the Bohr radius of the 
GaAs system and we use this as our length unit throughout this paper. In
Fig. \ref{fig0} we use the parameters: $m = 0.067 m_0$
and the wire width $a = a_B$. As discussed above, we find that the 
plasmon dispersion
strongly depends on the  interwire Coulomb
correlation determined by $d$. 
In these figures we find that for larger spatial separation the energy 
difference of $\omega_{\pm}$ modes is small
because as $d$ increases the interwire Coulomb correlation decreases.
The important point to note is that 
both the $\omega_{\pm}$ modes remain undamped upto very large wave 
vectors due to the suppression of 1D SPE, and therefore experimental 
observation of $\omega_{\pm}$ charge density modes in coupled bi-wire 
systems should be relatively easy via light scattering or 
far-infrared spectroscopy.

Now we consider the collective mode spectrum in a general system where the two
components may have different densities and masses.
The general situation with $d=0$ is of special experimental relevance 
because it describes the photoexcited 1D EHP system --- we therefore 
defer the discussion of the $d=0$ 1D EHP system to section IV.
Since there is a low energy gap in the
1D SPE, we look for undamped plasmons in
two regions: the high frequency region ($\omega > qv_{F1}, \;
qv_{F2}$) and the low-frequency region ($qv_{F1} > \omega >
qv_{F2}$). (We choose $v_{F1} > v_{F2}$ throughout this section without
any loss of generality.) 
In the long wavelength limit ($q \rightarrow 0$), 
we solve Eq. (\ref{zeros}) using Eq. (\ref{fix}) to get
\begin{equation}
\omega_{+}(q)  =  q v_{F1} \left [ 1 + \frac {2(1+\alpha)} {\pi v_{F1}}
V_1(q\rightarrow 0) \right ]^{1/2},
\label{wp}
\end{equation}
\begin{equation} 
\omega_{-}(q)  =  q v_{F1} \left [ \alpha + \frac{8}{\pi v_{F1}}
\frac{\alpha}{1 + \alpha}V_-(q\rightarrow 0) \right ]^{1/2},
\label{wm}
\end{equation} 
where $\alpha=v_{F2}/v_{F1} = (n_2/n_1) (m_1/m_2) \le 1$.
In the long wavelength limit, these asymptotic expressions for
$\omega_{\pm}$ are valid in both the high frequency
($\omega > qv_{F1}$) and the low frequency ($qv_{Fl} > \omega > q
v_{F2}$) regimes.
The condition for the existence
of an undamped AP mode ($\omega_-$) in the high frequency
regime($\omega > qv_{F1}$)  is
given by $v_-(\alpha,d)>v_{F1}$, or
\begin{equation}
\alpha + \frac{8}{\pi v_{F1}}
\frac{\alpha}{1 + \alpha}V_-(q\rightarrow 0)  > 1.
\label{cond}
\end{equation}
We find that the condition defined by the inequality [Eq. (\ref{cond})]
leads to the conclusion that within RPA  $\omega_-$ can exist 
as an undamped mode either for
a large spatial separation (and arbitrary $\alpha$) or for almost 
equal Fermi velocities, i.e., $\alpha \approx 1$, (and arbitrary $d$).
As $\alpha$ decreases the system should have a large spatial separation
to have an undamped high energy AP mode. For
$\alpha=1$ (e.g., equal mass and density), we find that the acoustic mode 
lies  in the high
frequency regime and is always undamped for arbitrary $d$. 
This is also true in 2D bilayer systems with finite layer
separations, where the AP mode is undamped  as long as
$\alpha=1$ and $d \neq 0$.
The condition for the existence
of the AP mode  in the low frequency
regime ($qv_{F1} > \omega_- > q v_{F2}$) is
given by $ v_{F2} < v_-(\alpha,d) < v_{F1}$.
Due to the energy gap in the 1D SPE the AP mode is undamped in this regime.
The existence of the undamped AP mode in the low frequency regime for
arbitrary $\alpha$ and $d$ is a unique phenomenon in the 1D two
component system arising entirely from the non-existence of low energy 
SPE in 1D. In 2D or 3D two-component system the low frequency regime
is necessarily accompanied by Landau damping due to the presence of 
the low energy SPE continua.

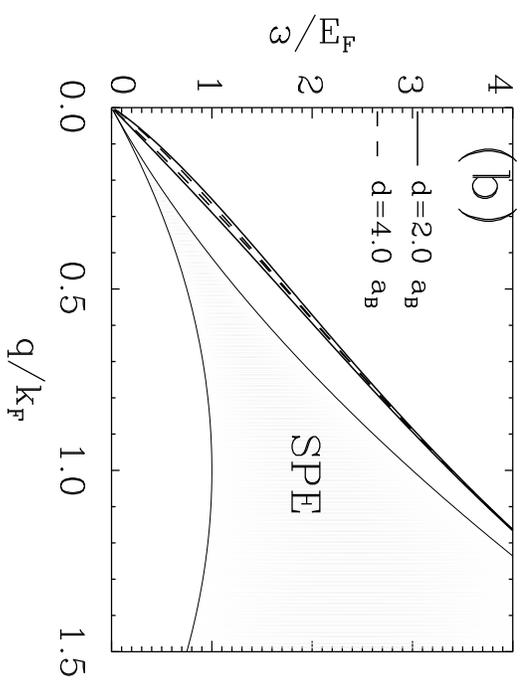
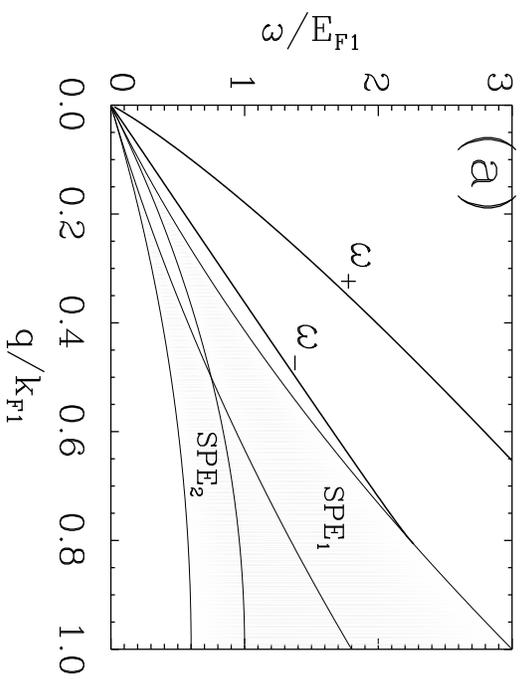
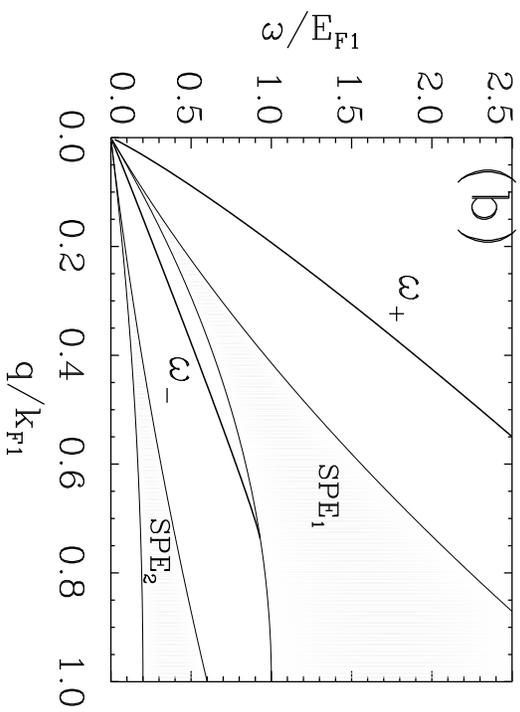
\begin{figure}
\caption{ 
Calculated RPA collective mode dispersion   (a) for the
spatial separation $d=0.5 a_B$ and $\alpha = v_{F2}/v_{F1}= 0.6;$ 
and (b) for $d=0.2 a_B$ and $\alpha = 0.2$. 
The $\omega_+$ ($\omega_-$) lines denote the OP (AP) mode and
SPE$_1$ (SPE$_2$) denote the single particle excitation of the
component 1 (2).
}
\label{fig1}
\end{figure}

In Fig. \ref{fig1} we show our numerically calculated longitudinal collective
modes (for different $d$ and $\alpha$) with a fixed density 
$n_1 = 0.6 \times 10^6 \; {\rm cm}^{-1}$, effective 
mass $m_1 = 0.067 m_e$ (corresponding to the effective
electron mass of the GaAs), and a wire confinement $a=a_B$.
In these figures the higher (lower) 
energy mode $\omega_+$  ($\omega_-$) denotes the optical (acoustic) mode.
In Fig. \ref{fig1}(a)  we use the
parameters: the spatial separation of the two wires $d=0.5 a_B$ and
$\alpha = (n_2/n_1)(m_1/m_2) = 0.6$. (Note that $\alpha$ determines the
density of the component two for a given $n_1$ and $m_1$, $m_2$. 
For example, if $m_2 = m_1$ we have $n_2= 0.6 n_1$.) 
With these parameters the
velocity of the acoustic mode $v_-$ exceeds the Fermi velocity of the 
component one ($v_- > v_{F1}$) and lies well above the SPEs of both
components (high frequency regime).
In Fig. \ref{fig1}(b) we choose the parameters which limit the $v_-$
between $v_{F1}$ and $v_{F2}$: $d=0.2 a_B$ and $\alpha = 0.2$. (Note that
this requirement restricts $d$ to unrealistically small values.)
Thus, the undamped AP mode with these conditions lies in low frequency
regime ($v_{F2} < v_- <v_{F1}$)  which corresponds to the regime between
two  SPEs.  The important point is that in both cases (``large'' $d$ 
and $\alpha$, and ``small'' $d$ and $\alpha$) the AP is undamped 
in 1D in contrast to higher dimensional systems.

Next, we investigate the plasmon modes of the two-component
system in the presence of  impurity scattering. 
We calculate the plasmon dispersion of a spatially separated 1D
two-component system with equal masses and densities ($\alpha=1$)
in the presence of impurity scattering (characterized by a constant 
level broadening $\gamma$).  With the approximation for the 
polarizability function
for small $q$ and $\omega$ defined by
Eq. (\ref{piga}) the plasmon dispersion is given by
\begin{equation}
\omega_{\pm}(q) \sim -i\frac{\gamma}{2} + \sqrt{-\frac{\gamma^2}{4} +
\frac{2n q^2}{m} V_{\pm}(q)},
\label{wgam}
\end{equation}
where $\omega_+$ ($\omega_-$) indicates the OP (AP) mode.
Note that for
$\gamma \rightarrow 0$, Eq. (\ref{wgam}) recovers the result
previously obtained for the pure two-component system.
Thus, in the presence of impurity scattering  the
plasmon modes acquire a wave 
vector dependent imaginary part which corresponds to a plasma damping,
$\omega_{\pm}(q) = \omega_{p\pm}(q) - i \alpha_{\pm}(q)$. The plasmon 
becomes overdamped at small $q$ as it 
becomes completely imaginary for small $q$, i.e.,  the
plasmon modes are totally overdamped below the critical wave vectors
$q_{c\pm}$ and  exist only for wave vectors $q$ larger than these critical
values. 
(Note that the plasmon broadening $\alpha_{\pm}(q)$ is, in
general, determined by the imaginary part of the complex plasmon mode,
and is different from the level broadening $\gamma$.)
For $q > q_{c\pm}$ the plasma dampings are roughly
comparable to the half of the impurity scattering, i.e.,
$\alpha_{\pm}(q) \approx \gamma/2$. From
Eq. (\ref{wgam}) we find the critical wave vectors
(below which the plasmon is no longer a well defined collective mode) as 
\begin{eqnarray}
q_{c+} & = & \frac{K}{\sqrt{2}|\ln(Ka/\sqrt{2})|}, \nonumber \\
q_{c-} & = & \frac{K}{\sqrt{\frac{\epsilon_b}{2e^2}V_-(q=0)}},
\label{qcpm}
\end{eqnarray}
where $K=\gamma \sqrt{m\epsilon_b/(8ne^2)}$.  As the impurity
scattering rate $\gamma$ increases the critical wave vectors $q_{c\pm}$
also increase. The $q_{c+}$ (just as $\omega_+$ itself) depends weakly 
on $d$, but $q_{c-}$ (just as $\omega_-$) is strongly affected by $d$.
As $d \rightarrow 0$
the intra- and
inter-wire Coulomb interactions become the same, $V_-(q) \rightarrow 0$, 
and consequently $q_{c-}
\rightarrow \infty$. Thus, for the system with a large impurity
scattering and  a small spatial separation we have only the OP mode in
the long wave length limit. 
In general, we  have $q_{c-} > q_{c+}$ from Eq. (\ref{qcpm}), which
means that the AP mode is easily affected by impurity
scattering (and is therefore more difficult to observe experimentally.)

In Fig. \ref{fig2} we show the plasmon dispersion calculated numerically by
using the full polarizability, Eq. (\ref{pig}), in the presence of
finite impurity scattering.
We find zeros of the complex dielectric function
of the 1D two-component system in the presence of impurity
scattering,  i.e., $\epsilon(q,\omega) = 0$, to obtain the 
plasmon dispersion curves.  In Fig. \ref{fig2}
the curves with $\omega > 0 $ give the plasma 
frequency or the real part
[$\omega_p(q)$], and those with $\omega<0$ give the plasma damping or
the imaginary part ($\alpha$) of the complex zero solutions
[$\omega(q)$]. 
The figure shows that the plasmons are 
overdamped below critical wave vectors $q <
q_{c\pm}$. The critical wave vectors, $q_{c\pm}$, below which the plasmon does
not exist due to impurity scattering effects, depend on the density and the
impurity scattering rate. We note that the overdamping of the plasmon mode
occurring for 
small $q$ is a direct consequence of the diffusive nature of the
electron dynamics. We emphasize that this long wavelength impurity 
scattering-induced suppression of collective modes is more severe 
in 1D than in 2D or 3D (in 3D, because the OP is gapped, it is hardly 
affected by impurity scattering.)

Finally, we study the dynamical structure factor, $S(q,\omega) \propto {\rm
Im}[\epsilon(q,\omega)^{-1}]$,  which gives a direct measure of the 
oscillator strength of the density 
fluctuation spectrum, and many experimental probes such as
inelastic electron  
and Raman scattering spectroscopies are directly related to the dynamical
structure factor \cite{mahan}. Equivalently, the dynamical structure
factor is a measure of the spectral weight
carried by the particular collective mode of the electron
liquid. An undamped plasmon shows up as a $\delta$-function peak in
$S(q,\omega)$ indicating the existence of a simple zero of
$\epsilon(q,\omega)$. When both Re$[\epsilon]$ and
Im$[\epsilon]$ become zero [i.e., $\epsilon(q,\omega) = 0$]
the imaginary part of the inverse dielectric function
Im$[\epsilon(q,\omega)^{-1}]$ is a
$\delta$-function with the strength 
\begin{equation}
W(q)=\frac{\pi} {|\partial {\rm Re}[\epsilon(q,\omega)] /
\partial \omega|_{\omega=\omega_{\pm}(q)}},
\end{equation}
where $\omega_{\pm}(q)$ are the plasmon dispersions for the system.
The damped plasmon ($\alpha \neq 0$), however,
corresponds  to a broadened peak in $S(q,\omega)$ -- for larger
broadening, as for $q < q_c$  in the system with impurity scattering,
the plasmon is overdamped and there is no peak in $S(q,\omega)$.
At long wavelengths ($q \rightarrow 0$), the optical plasmon $\omega_+$
mostly exhausts the $f$-sum rule and carries almost all the spectral weight. 
$W(q)$ approaches zero as $q\rightarrow 0$ because the 1D plasma
frequency vanishes.  
Note that the small weight of the AP mode at long wavelengths
makes it particularly susceptible to collisional damping effects in 1D
two component system.
For the system with the same effective mass  and a large spatial
separation, the acoustic plasmon mode has an energy comparable to that
of the optical plasmon mode, which gives the enhanced spectral weight
associated with the AP mode. 

In Fig. \ref{fig3} the Im$[\epsilon(q,\omega)^{-1}]$ is
plotted for different wave vectors
as a function of the frequency for two different systems.
In Fig. \ref{fig3}(a) we choose the parameters such that the AP
mode lies in the low frequency regime: $n_1= 1.0\times10^6
cm^{-1}$, $a=2 a_B$, 
$\alpha=0.2$, and $d=0.5a_B$. In Fig. \ref{fig3}(b) we choose the
parameters such that the AP 
mode lies in the high frequency regime: $n_1= 0.6\times10^6
cm^{-1}$, $a = a_B $, $\alpha = 0.9$, and $d=a_B$.  
The insets in Fig. \ref{fig3} show the weight $W(q)$ of the plasmon
modes.
For pure system the plasmon peaks become  $\delta$-function
peaks with  weights given in the inset, but with impurity scattering rate
$\gamma_1 = 0.1 E_{f1}$ and $\gamma_2 = 0.1 E_{f2}$, we show broadened
peaks giving the damped plasmon modes.  
As shown in Fig. \ref{fig3} the spectral weight of the acoustic mode is
comparable with that of the optical mode at finite wave vectors,
but not in the long wavelength limit. With the enhanced spectral
weight the acoustic plasmon mode in 1D should be
more easily observable than in higher
dimensions and may significantly affect  physical properties 
of the quasi-1D two-component systems.
The results presented in this section indicate that the AP mode 
should be observable in double wire structures under experimentally
realizable conditions.

\section{Bi-wire with weak tunneling}

In this section we analytically calculate the collective mode dispersions in
bi-wire structures ($\alpha = 1$, i.e., $n_1 = n_2$ and $m_1 = m_2$,
and $d \neq 0$)
{\it in the presence of significant interwire 
quantum tunneling}, all earlier work in the previous section having
dealt with  the zero tunneling limit. 
Since our focus is on understanding tunneling effects in this section, 
we put $\alpha =1$ with no loss of generality.
In the presence of interwire tunneling, the 
electron energy eigenstates $E_{\pm}$ should be used \cite{dassarma1}
as the basis set 
rather than the wire index which is no longer a good quantum number.
The energy levels
$E_{\pm} = \varepsilon(k) \pm t$, where $\varepsilon(k) = k^2/2m$ is
the parabolic one electron 1D kinetic energy in each wire and $t$ is the
tunneling strength, are  the usual symmetric and
antisymmetric one electron eigenstates in the presence 
of tunneling with the single
particle symmetric-antisymmetric (SAS) gap given by $\Delta_{SAS} =
E_+ - E_- = 2t$. In the SAS representation the collective mode spectra
become decoupled by virtue of the symmetric nature of our bi-wire
system (i.e., both wires identical with equal electron
density), and the collective density fluctuation spectra are given by
the following two equations for the in-phase and the out-phase plasmon
modes $\omega_{\pm}$ respectively:
\begin{equation}
\epsilon_{+}(q,\omega) = 1-V_+(q) \left [ \Pi_{++}(q,\omega) +
\Pi_{--}(q,\omega) \right ] = 0,
\label{ep}
\end{equation}
and
\begin{equation}
\epsilon_{-}(q,\omega) = 1-V_-(q) \left [ \Pi_{+-}(q,\omega) +
\Pi_{-+}(q,\omega) \right ] = 0,
\label{em}
\end{equation}
where $V_{\pm}(q) = V(q) \pm U(q)$ with
$V(q)$ and $U(q)$ being respectively the intrawire and interwire Coulomb
interaction matrix elements given in Eqs. (\ref{vq}) and
(\ref{uq}). The $\Pi_{\alpha \beta}(q,\omega)$ in
Eqs. (\ref{ep}) and (\ref{em}), with $(\alpha,\beta)=(+,-)$, are the
noninteracting SAS polarizability functions within our RPA theory:
\begin{equation}
\Pi_{\alpha \beta}(q,\omega) = 2\int \frac{d^2k}{(2\pi)^2}
\frac{f_{\alpha}({\bf k}+{\bf q}) - f_{\beta}({\bf k})}{w +
E_{\alpha}({\bf k}+{\bf q}) - E_{\beta}({\bf k})},
\label{piab}
\end{equation}
where $f_{\alpha,\beta}$ are Fermi occupancy factors.

Solving Eqs. (\ref{ep})--(\ref{piab}) we obtain the collective density
fluctuation spectra of the coupled bi-wire system. In the absence of
tunneling, $t=0$, one has $E_+=E_-=\varepsilon({\bf k})$, and one then
recovers in a straightforward fashion the results we have in Sec. II, i.e.,
the in-phase optical 
($\omega_+ \sim q |\ln(qa)|^{1/2}$) and the out-of-phase 
acoustic ($\omega_- \sim q$) plasmons of 
a bi-wire system without any electron tunneling. It is, in fact,
straightforward  to obtain analytically [from
Eqs. (\ref{ep})--(\ref{piab})] the long
wavelength ($q \rightarrow 0$) plasma modes of the coupled bi-wire
system {\it including effects of interwire tunneling}. We obtain in the
long wavelength limit the following results:
\begin{equation}
\omega_{+}^2(q \rightarrow 0) = q^2 \left [ v_F^2 + \frac{2}{\pi}
(b_+ + b_-) v_F V_+(q) \right ],
\label{wpi}
\end{equation}
\begin{equation}
\omega_-^2(q \rightarrow 0) = \Delta_{SAS}^2 +
\frac{4k_F}{\pi}(b_ + -b_-)V_-(q=0)\Delta_{SAS}, 
\label{wmi}
\end{equation}
where $v_{F} = k_F/m$ is the Fermi velocity and $b_{\pm} = (1 \pm
t/E_F)^{1/2}$ for $n > n_c = 4m \Delta_{SAS}/\pi^2$ when both
symmetric and antisymmetric levels are occupied (i.e., the 1D Fermi
energy $E_F > \Delta_{SAS}$), and $b_+=2,b_-=0$ for 
$n \leq  n_c$ when only the symmetric level is  occupied (i.e., $E_F <
\Delta_{SAS}$).

The most important qualitative feature of the plasmon dispersion in
the presence of interwire tunneling 
is that even though the in-phase OP mode ($\omega_+$)  depends weakly on
tunneling, the out-of-phase mode ($\omega_-$), which
is purely acoustic in the absence of tunneling ($\omega_- \sim q$ if
$\Delta_{SAS} = 0$), develops a plasmon gap at $q=0$ in the presence
of nonzero tunneling. 
The plasmon gap $\Delta \equiv \omega_-(q=0)$ 
depends nontrivially on the 1D electron density $n$, the interwire
tunneling amplitude $t$, and the interwire distance $d$. 
It is easy to see from Eq. (\ref{wmi}) that this
plasmon gap $\Delta$, has the following behavior
\begin{equation}
\Delta \sim \Delta_{SAS} \;\; {\rm or} \;\; \sqrt{\Delta_{SAS}},
\end{equation}
depending on whether the interwire tunneling is strong or weak. It 
should be emphasized that the strikingly non-intuitive $\Delta_{SAS}^{1/2}$
dependence of the collective mode gap (on the square root of the
single particle gap) is purely a Coulomb interaction effect, which
dominates the collective excitation spectra in the weak tunneling
situation. We mention the curious phenomenon that interwire tunneling 
converts the $\omega_-$ mode into an effective optical mode by opening
up a long wavelength plasmon gap (for $t=0$ the $\omega_-$ mode is the 
AP mode).

\section{photoexcited one dimensional electron-hole plasma}

In sections II and III we discussed a double wire system where the two
one dimensional charged components are separated by a distance
$d$. Now we consider the possibility of acoustic plasmons in a
photoexcited homogeneous (i.e., $d=0$) 1D EHP where
the electrons and holes are not spatially separated. This is, in fact,
the original context\cite{pines} in which the AP mode was first
discussed in the solid state plasma more than forty years ago. The 
photoexcited EHP in semiconductors is also the
system in which most of the early investigations \cite{pinczuk} of
acoustic plasmons 
were carried out. Our reason for singling out the 1D EHP for a 
detailed investigation in this section (in principle, it is just 
the $d=0$ limit of Sec. II) is the existence of 1D EHP in real 
samples \cite{weg} from several groups where a search for 1D 
$\omega_{\pm}$ modes via light scattering spectroscopy should be successful.

It has been known for a long time \cite{pines} that a two-component EHP
allows for the existence of two collective modes, one of which (the
low frequency one) is linear in wave vector or acoustic at long
wavelengths, and the other mode (the high frequency or the optical
plasmon) is essentially the combined collective charge density
oscillation of the whole system being therefore qualitatively similar
to the usual plasmon mode in a single component plasma. In an
electron-hole two-component plasma, as appropriate in a photoexcited
semiconductor system, the AP corresponds to the in-phase
collective density fluctuation excitation of electrons and holes, and
the OP corresponds to their out-of-phase motion with
respect to each other. [If the two components are both electrons or
both holes, for example, in a bi-wire (as considered in Sec. II of the
paper) or bilayer or a two-band system, then the AP
signifies the out-of-phase collective mode and the OP the
in-phase motion -- in general, the acoustic mode is the neutral mode
at long wavelengths and the optical mode the charged mode.] In spite
of compelling theoretical arguments \cite{pines,pines1,platzmann} for
the existence of ``quantum'' 
AP in degenerate two-component homogeneous
electron-hole solid state plasmas the experimental observation of low
temperature AP in two and three dimensional homogeneous
electron-hole semiconductor plasmas has been very severely hindered by
Landau damping effects. It turns out that in two and three dimensional
degenerate homogeneous EHP the AP is
necessarily Landau damped (by decaying into single particle
excitations) and is in fact unobservable as a well-defined mode unless
the effective mass ratio between the two species is extremely
large \cite{pines,pines1,platzmann}. This has prevented an unambiguous
observation \cite{pinczuk} of a quantum 
AP in degenerate solid state EHP
although the corresponding classical AP (in the
non-degenerate high temperature electron-ion plasma) was
experimentally observed \cite{alexeff} almost forty years ago.

In this section we show that the elusive quantum AP in
the homogeneous (i.e., $d=0$) EHP should be
relatively easy to observe in the degenerate 1D EHP \cite{weg}
confined in semiconductor quantum wire structures
due to the severe suppression of single particle excitations in one
dimensional systems. In particular, within  the RPA which should be
quantitatively valid in one dimension by virtue of the smallness of
vertex corrections, the AP mode in a 1D EHP
in GaAs quantum wires should be {\it completely} undamped upto
a rather large wave vector $q \ge k_F$, in contrast to two/three
dimensional EHP where the AP is damped
in GaAs and is essentially unobservable. We believe that resonant
inelastic light scattering spectroscopy, which has already been 
successfully used \cite{goni} in observing the 1D plasmon in {\it single
component} ($n$-doped) GaAs quantum wires, is the ideal tool in the
search for the AP in {\it photoexcited} (undoped) two
component EHP in GaAs quantum wires. Theoretical
results presented in this section show that the 1D in-phase AP
mode should be readily experimentally observable in resonant
inelastic light scattering experiments performed on photoexcited 1D
EHP \cite{weg} in narrow GaAs-AlGaAs quantum wires.

We put $d=0$ in the formalism developed in Sec. III of the paper and
allow for the possibility of different masses $m_e$ and $m_h$ taking
equal densities $n_e=n_h = n$ of the electrons and holes.
Below we present and discuss our results for collective
modes in 1D EHP both in the RPA and in the Hubbard approximation (HA)
where one includes a local field correction to the RPA non-interacting
polarizability. The inclusion of local field corrections (HA) beyond
RPA distinguishes this section from Sec. II where strict RPA was used ---
our motivation for using HA (as well as RPA) is to make our results 
more realistic for the experimental systems where local field 
corrections may very well be important at the experimentally 
feasible densities and wave vectors.
The dynamical dielectric function in the HA is given by
\begin{eqnarray}
\epsilon_H(q,\omega) & = &\left \{ 1 - V(q)[1-G_e(q)]\Pi_e^0(q,\omega)
\right \} \left \{ 1 - V(q)[1-G_h(q)]\Pi_h^0(q,\omega)
\right \} \nonumber \\ 
 &  & - V(q)^2\Pi_e^0(q,\omega)\Pi_h^0(q,\omega),
\label{pih}
\end{eqnarray}
where $G_{e,h}(q)$ is a  simple static local
field correction to the RPA and is given by $
G_{e,h}(q) = V(\sqrt{q^2+k_{Fe,h}^2})/2V(q)$, where $k_{Fe,h}$ are the
Fermi wave vector of the electron and hole respectively.
The calculation in the HA is similar to that in the RPA with the RPA
dielectric function [Eq. (\ref{epsi})] being replaced by the HA
dielectric function [Eq. (\ref{pih})].
By putting $d=0$ we have $V_-(q) = 0$, $V_+(q) = V(q)$.  Thus, in the
long wavelength limit $q\rightarrow 0$,  we have the plasmon mode
dispersions within RPA from Eqs, (\ref{wp}) and (\ref{wm})
\begin{equation}
\omega_{+}(q)  =  q v_{Fe} \left [ 1 + \frac {2(1+\alpha)} {\pi v_{Fe}}
V(q\rightarrow 0) \right ]^{1/2},
\label{wph}
\end{equation}
\begin{equation} 
\omega_{-}(q)  =  \sqrt{\alpha} q v_{Fe},
\label{wmh}
\end{equation} 
where $\alpha\equiv v_{Fh}/v_{Fe}$ corresponds to the mass ratio 
$m_e/m_h$ since $n_e = n_h$.
As the mass ratio decreases ($\alpha \rightarrow 0$), the OP mode ($\omega_+$)
becomes the plasmon mode of the one component system. Since $0<\alpha <
1$ we have $q v_{Fh} < \omega_-(q) < qv_{Fe}$. The AP mode
within RPA lies always 
in the undamped region (low frequency regime) between two SPEs. This
undamped 1D AP mode $\omega_-$ in the long wavelength limit
is fundamentally different from the AP mode of the higher dimensional 
EHP since
the presence of the low energy Landau damping region in 2D or 3D gives
rise to the complete damping of AP mode, unless $\alpha$ is very small --- in
the 1D EHP the AP mode is undamped within RPA for arbitrary values of 
$\alpha$ ($\alpha < 1$ by choice).

In Figs. \ref{fig4} --\ref{fig6} we show the numerically calculated
plasmon modes and spectral weights for different densities in the
photoexcited GaAs
quantum wire 1D EHP within both
RPA and HA. In Figs. \ref{fig4} --\ref{fig6} we use the parameters
corresponding to GaAs: 
$m_e = 0.067 m_0$, $m_h = 0.45 m_0$, $\epsilon_b = 11$, and the
confinement width of the wire $a = a_B$. We also use a level
broadening $\gamma_{e,h} = 0.1E_{Fe,h}$ due to the impurity scattering
in calculating the mode spectral weights.
Note that in the long wavelength limit the RPA AP modes are
undamped (except, of course, for impurity damping) for all 
densities because of the absence of long wavelength
Landau damping.  
In general, similar to the corresponding higher dimensional situations
the local field effects in the HA reduce the plasma
frequencies \cite{dashwang} compared with the RPA results, 
in particular, in the low density
system. Since the $\omega_+$ mode exhausts the $f$-sum rule 
in the long wavelength limit, 
the local field correction has relatively little effect on the 
$\omega_+$ mode. The $\omega_-$ AP mode is, however, very 
strongly affected by local field effects.
Fig. \ref{fig4} shows the results for the EHP density $n_e = n_h = 10^{6} \;
{\rm cm}^{-1}$: At this density we find the remarkable result that the 
$\omega_-$ AP mode is completely suppressed by local field effects
as it gets pushed into the SPE Landau damping continua. Thus,
in Fig. \ref{fig4}(a) the AP mode dispersion within HA  does
not show up because $\omega_-$ lies completely in the Landau damping 
continuum. Fig. \ref{fig4}(b) shows that within RPA (solid lines) the 
$\omega_-$ mode (low 
energy peaks) carries an appreciable spectral weight 
which is comparable to that carried by  the
OP mode $\omega_+$ (high energy peaks), however,  within HA
(dashed lines) the  $\omega_+$ mode actually carries all the spectral 
weight (we find only
the high energy $\omega_+$ peaks in the HA with the low energy $\omega_-$
peaks being completely suppressed.)
As density increases the (completely suppressed) $\omega_-$ mode in HA
reappears in the low energy regime as the AP mode comes out of the 
Landau continuum.
In Figs. \ref{fig5} and \ref{fig6} we show the results for density
$n_e = n_h = 1.2 \times 10^{6} \; {\rm cm}^{-1}$ and $n_e = n_h = 1.5
\times 10^{6} \; {\rm cm}^{-1}$, respectively.  At the density  $n_e = n_h
= 1.2 \times 10^{6} \; {\rm cm}^{-1}$ the $\omega_-$ mode
reappears below the SPE$_h$ in the HA. 
This $\omega_-$ mode lying below SPE$_h$ is undamped due
to the low energy SPE gap in 1D and is a characteristic feature of 1D EHP.
The spectral weight of the undamped
$\omega_-$ mode in the low energy region (at higher density) is 
comparable to that of the $\omega_+$ mode even
within  HA [Fig. \ref{fig5}(b)]. At very high density
(Fig. \ref{fig6}) $n_e = n_h = 1.5
\times 10^{6} \; {\rm cm}^{-1}$ the local field corrections are
weak, and we find that the  $\omega_-$ mode (in the HA) lies 
between SPE$_e$ and SPE$_h$ and is undamped in the long wave length
limit.  The calculated spectral weight in the HA is comparable
to that in the RPA in these high density systems.
One conclusion of our HA-based calculations is that one should 
use higher EHP densities for an unambiguous observation of 
undamped 1D AP --- this is consistent with our RPA results 
also, where Landau damping is absent, because in RPA higher 
carrier densities imply quantitatively weaker effects of 
impurity level broadening.

Given the unique nature of the SPE continua (and the strong low 
energy suppression of Landau damping) in the 1D system, the question 
naturally arises about whether the collective mode behavior in 1D EHP 
is fundamentally different from that in 2D and 3D systems, where the 
AP mode is usually unobservable due to its unavoidable decay via 
Landau damping to the electron(i.e., the lighter mass component) 
SPE unless $\alpha$ is very small (i.e., the mass ratio very small, 
$m_e \ll m_h$) which cause the Landau damping to be small, but 
still non-zero. In the 1D EHP, however, there are three (as opposed to 
only one in 2D and 3D EHP) Landau damping-free regimes in the 
$\omega$-$q$ space --- these are [see Fig. 5(a) or 6(a) or 7(a)] the 
high energy-low wave vector regime above the lighter mass SPE continuum 
(which is present in 2D and 3D systems also), the low energy regime 
below both the SPE continua, and the intermediate energy --- low to 
intermediate wave vector regime in the SPE gap between the SPE 
continua of the two components (we emphasize that Landau damping 
of collective modes is allowed in 2D and 3D systems in the last 
two regimes even in the RPA because there is no low energy gap 
in the SPE continuum in higher dimensions). To emphasize the 
qualitative difference between 1D EHP and its higher dimensional 
counterparts with respect to the Landau damping characteristics of 
the collective modes, we provide in Figs. 8 and 9 our calculated RPA 
results for collective mode dispersion and damping in 2D and 3D EHP 
respectively, choosing the same effective mass, lattice dielectric 
constant, level broadening, and ``approximately equivalent'' carrier 
densities. One can see that in Figs. 8(a), the 2D EHP case, and 9(a), 
the 3D EHP case, the OP mode is above the SPE continua of both 
components (exactly the same as in Figs. 5 -- 7 for the OP in 1D EHP) 
and is therefore not Landau damped whereas the AP mode in both case 
lies within the electron (i.e., the lighter mass) SPE continuum (but 
above the hole SPE) and is therefore Landau damped by decaying into 
the quasiparticles of the electron component in the EHP. The contrast 
between the RPA results for the 2D/3D AP mode (Figs. 8 and 9) and the 
1D AP mode (Figs. 5-7) is striking in terms of Landau damping --- the 
2D/3D AP mode is invariably Landau damped by being inside the SPE 
continuum of the lighter carrier component (i.e., the higher Fermi 
velocity, $v_{Fe}$, in our case) where the 1D AP mode (within RPA) 
is undamped by being in the intermediate regime in between (the SPE 
``gap'' regime) the SPE continua of the two components and is 
therefore {\it not} Landau damped by either plasma component. This 
non-existence of AP Landau damping in the SPE ``gap'' regime in a 
1D EHP is a fundamental one dimensional effect arising from the 
suppression of long wavelength single particle excitations peculiar 
to 1D. We should mention, however, that, as is obvious from Fig. 5-7, 
inclusion of local field corrections within the HA may push the AP mode 
down into the SPE continua of the higher mass plasma component, i.e., 
the hole SPE, making the AP mode strongly Landau damped even in the 1D 
EHP --- our comparison between 1D and 2D/3D EHP collective modes and 
their qualitative difference are based on the RPA, going beyond RPA 
may very well render the 1D AP unobservably Landau damped as happens, 
for example, within the HA in our calculation for $n < 1.2 \times 
10^6$ cm$^{-1}$ in Figs. 5 and 6.

It is very anticlimactic, however, that even within the RPA the 
qualitative difference in the Landau damping behavior of the AP in 
the 1D EHP compared with its higher dimensional counterparts is not 
dramatically reflected in the mode spectral weight or the dynamical 
structure factor, as can be seen by comparing the RPA spectral weights 
in Fig. 5(b), 6(b), 7(b) in the 1D EHP with those in the 2D/3D EHP 
given in Figs. 8(b)/9(b). (We emphasize that we use equivalent impurity 
level broadening, $\gamma = 0.1 E_F$, in all our spectral weight 
calculations.) In general, the spectral weight carries by the AP mode 
(the lower frequency peak in the dynamical structure factor 
Im[$\epsilon^{-1}$] for each $q$) is comparable, albeit slightly 
larger, in the 1D EHP as that in the 2D/3D EHP. In particular, the 
AP spectral weights in the 1D and the 2D EHP seem to be quantitatively 
similar whereas the corresponding 3D AP mode carries considerably less 
spectral weight (which is understandable based on the fact that in 3D 
the OP has a finite gap at long wavelengths in contrast with the 1D/2D 
OP which vanishes at long wavelengths, and therefore in a 3D EHP the 
long wavelength OP mode carries considerably more relative spectral 
weight than its lower dimensional counterparts). 
The fact that the 1D AP carries small spectral weight in spite of its 
lack of Landau damping can be easily understood on the basis of the 
$f$-sum rule, which is essentially completely exhausted by the OP in 
the long wavelength limit, leaving rather little spectral weight to 
be carried by the 1D AP in the long wavelength. At large wave vectors 
(i.e., away from the long wavelength limit), on the other hand, the AP 
is Landau damped (because it enters the SPE continua) and all collective 
modes start carrying little spectral weights.

We therefore conclude this section on 1D EHP collective modes with two
relatively modest conclusions: (1) In spite of the non-existence 
of AP Landau damping in the 1D EHP within RPA, the calculated AP 
spectral weight is not much larger than the corresponding 2D situation 
(it is, however, substantially larger than the corresponding 3D situation), 
and therefore the observability of 1D EHP AP should be only slightly 
more favorable than the corresponding 2D case; (2) inclusion of local 
field corrections in general severely suppresses the AP spectral weight 
in the 1D EHP, and at least within the HA the AP becomes unobservable 
at lower densities.

\section{two-component luttinger liquid}

In this section we briefly discuss, for  the sake of completeness, the AP mode 
of the two-component Luttinger liquid, which has  two different
Fermi velocities $v_{Fe}$ and $v_{Fh}$ corresponding to the electron
and the hole respectively. We make the standard Luttinger model approximation
of linearizing the single particle energy.
This system  can be described by the 2-component Luttinger model \cite{hal} 
Hamiltonian 
\begin{equation}
H = \sum_{i,k,\sigma}v_{Fi}k \left [ a_{i,k,\sigma}^{\dagger}
a_{i,k,\sigma} - b_{i,k,\sigma}^{\dagger}
b_{i,k,\sigma} \right ] + \frac{1}{2L} \sum_{i,j,q} V_{i,j}(q)
\rho_{i}(q)\rho_j(-q), 
\label{hl}
\end{equation}
where {\it i,j}={\it e,h}, and $a_{i,k,\sigma}$ ($b_{i,k,\sigma}$) is the
destruction operator of the right- (left-) moving $i$-th component
with momentum $k$ and spin $\sigma$, $V_{i,j}(q)$ the Coulomb interaction
between $i$-th and $j$-th component (here $V_{i,j}= V_{ee}=V_{hh}=-V_{eh}$, 
and $\rho_i(q) = \rho_{i,a}(q) +
\rho_{i,b}(q)$ the total density operator of the $i$-th component.

By standard bosonization methods \cite{mahan,hal} we can diagonalize 
Eq. (\ref{hl}) and find two eigenmodes corresponding to the charge density
excitations
\begin{eqnarray}
\omega_{\pm}^2(q) & = & \frac{q^2v_{Fe}^2}{2} \left [ 1 + \alpha^2 +
(1+\alpha)\tilde{V}(q) \right ] \nonumber \\
& \pm & \frac{q^2v_{Fe}^2}{2} \left \{ \left [ 1 + \alpha^2 +
(1+\alpha)\tilde{V}(q) \right ]^2 - 4\alpha \left [
\alpha + (1+\alpha)\tilde{V}(q) \right ] \right
\}^{1/2},
\label{wpml}
\end{eqnarray}
where $\alpha =v_{Fh}/v_{Fe}$ and $\tilde{V}(q) = 2 V_{ee}(q)/(\pi
v_{Fe})$; note that $V_{ee}=V_{hh}=-V_{eh}$ in the 1D EHP. 
In the long wavelength limit,
$q \rightarrow 0$,  Eq. (\ref{wpml}) becomes
\begin{equation}
\omega_+(q) = v_{Fe} q \left [ 1 + (1 +
\alpha)\tilde{V}(q) \right ]^{1/2},
\label{wpl}
\end{equation}
\begin{equation}
\omega_-(q) = \sqrt{\alpha}qv_{Fe}.
\label{wml}
\end{equation}
These long wavelength charge density modes defined by Eqs. (\ref{wpl}) and
(\ref{wml}) are exactly the same as the long wavelength 
collective modes in the 1D EHP within RPA [Eqs. (\ref{wph}) and
(\ref{wmh})]. This can be explained by the fact that  vertex
corrections to the irreducible polarizability of the 1D system vanish
\cite{dl} in the Luttinger model, and therefore RPA and Luttinger 
model results become identical in the long wavelength limit. Even 
though the charge density mode dispersions of the
Luttinger model are identical to the RPA collective mode 
predictions based on the 
Fermi liquid theory, the damping properties of the $\omega_-$ AP mode 
in the Luttinger model is strikingly different form the RPA result. In the
previous section we found that within RPA the AP mode in
the 1D EHP exists only at long wavelengths as  the AP
mode gets severely damped by Landau damping at finite wave
vectors where it enters  the electron single particle continua 
and is Landau damped.
However, since there are no equivalent SPE Landau damping
mechanisms in the Luttinger model the
AP mode  in the Luttinger model seems  not to be damped out even at the
finite wave vectors, and seems to exist at arbitrary wave vectors.

The above-discussed difference between the RPA (Fermi liquid) result 
and the Luttinger model (non-Fermi liquid) result seems to suggest a 
direct experimental approach to observe the predicted non-Fermi 
liquid-like behavior of a one dimensional system by studying its 
plasmon damping behavior which, according to the RPA (Luttinger) 
model, should (should not) manifest finite wave vector Landau damping. 
This is, however, misleading because the strict non-existence of any 
damping of the collective modes at any wave vectors is purely a result 
of the Luttinger {\it model} (rather than being a generic Luttinger 
{\it liquid} result), arising entirely from the linearization of the 
single particle energy [the first term in Eq. (\ref{hl})] in the 
Hamiltonian. In the generic Luttinger liquid behavior, where one does 
{\it not} linearize the single particle energy, higher order processes 
(which asymptotically vanish at low energies and long wavelengths) 
produce collective mode damping arising from the (multiboson) boson-boson 
collision processes within the bosonized theoretical description. Such 
multiboson collision processes, which are akin to phonon-phonon 
anharmonic scattering processes producing phonon decay beyond the 
harmonic crystal theory, are {\it irrelevant} in the renormalization 
group sense at long wavelengths, but do produce damping of the 
collective modes at finite wave vectors. In this sense, therefore, 
no fundamental distinction exists between the two theoretical approaches 
since both predict damping of the collective modes at finite wave 
vectors, but {\it not} at zero wave vectors.
One expects the AP damping in the Luttinger liquid theory to be a 
smooth function of wave vector $q$, vanishing only in the long 
wavelength limit $q \rightarrow 0$. This seems to be drastically 
different from our RPA damping result where the AP is undamped 
upto a wave vector $q_c$ when it becomes damped as it enters the 
SPE continuum. Going beyond RPA, however, one finds that the AP 
is damped even outside the SPE Landau damping continua due to 
multipair (i.e., multiple electron-hole excitations) productions, 
and therefore there may {\it not} be much qualitative difference 
between RPA and Luttinger liquid theories even for the AP damping 
properties. This issue, however, merits further investigations, 
and detailed theoretical calculations of the AP damping properties 
in the two-component Luttinger liquid may very well point to some 
significant differences with our RPA results, which then may lead 
to a direct method of differentiating between Luttinger liquid and 
RPA results, based on the experimentally measured damping properties. 
Very recently, such a calculation \cite{sam} has been carried out for 
a {\it one component} clean Luttinger liquid using the {\it short range} 
interaction which, however, has to be generalized to the realistic 
long-range Coulomb interaction in 1D quantum wires before any firm 
conclusion on the damping behavior contrast between RPA and Luttinger 
liquid may be reached. We emphasize that the AP damping property being 
manifestly a non-universal property (determined by irrelevant operators 
such as the band curvature), its actual quantitative calculation in the 
Luttinger liquid theory is problematic. 

\section{summary}

We have obtained the dispersion and spectral weight of collective 
charge density excitations in zero temperature two-component one 
dimensional quantum plasma confined in semiconductor quantum wire structures. 
In some sense this paper is a two-component generalization of our 
earlier work \cite{dashwang} where we studied the zero temperature 
collective charge density excitations of the one component 1D system 
in some details. In a separate earlier publication \cite{hwang} we 
consider a specific example of collective modes in a two-component 
1D electron system (namely, a two-subband 1D system) in the context 
of obtaining quantitative agreement with the experimental results of 
ref. \onlinecite{goni}. In the current paper we refrain from repeating 
any results from these earlier publications \cite{hwang,dashwang} of 
ours, and the current paper, along with refs. \onlinecite{hwang} and 
\onlinecite{dashwang}, forms a reasonably complete quantitative 
description of possible low lying collective charge density excitations 
in 1D semiconductor quantum wire structures.

Our main results in this paper are the following: (1) there are two 
possible collective charge density excitations in a two-component 
1D quantum plasma associated with the relative phase fluctuations 
in the two charge densities --- the OP mode is the in-phase 
(out-of-phase) mode when the two components are the same (opposite) 
kind (both electrons or both holes versus one component electron 
and the other component hole) and the AP is the out-of-phase (in-phase) 
mode in the same situation (in general, the AP is essentially a neutral 
excitation at long wavelengths and the OP is the collective oscillation 
of the total charge density); (2) in contrast to higher dimensional 
systems where the AP mode is often Landau damped because it lies in 
the SPE continuum of the faster moving charge component, the 1D AP 
mode is invariably undamped at long wavelengths (within RPA) due to 
the severe suppression of long wavelength SPE continua in 1D; (3) this 
RPA result of the non-existence of long wavelength Landau damping in 
the 1D AP mode is drastically affected by local field corrections,
which at low density may overdamp the AP mode by pushing it inside 
the SPE continua of both components --- the OP mode on the other hand 
is relatively robust with respect to local field corrections since its 
long wavelength dispersion is fixed by the $f$-sum rule; (4) the AP 
mode in general carries little spectral weight at long wavelengths, 
and even at finite wave vectors its spectral weight is only slightly 
enhanced with respect to the corresponding 2D two-component situation, 
leading to the somewhat disappointing and unexpected conclusion that 
in spite of the complete suppression of the long wavelength SPE continua 
in 1D systems the neutral AP mode is not substantially easier to observe 
in 1D than in 2D/3D systems; (5) finite inter-wire separation in general 
enhances the AP spectral weight increasing its observability; (6) finite 
impurity scattering induced level broadening gives rise to a critical 
wave vector below which the collective modes are always overdamped, and 
thus the concept of a long wavelength collective mode does not strictly 
apply; (7) finite inter-wire tunneling produces a long wavelength plasma 
gap in the AP mode; (8) Luttinger liquid theory and Fermi liquid-based 
RPA theory produce formally identical OP and AP collective mode behavior 
in 1D two component systems making it impossible to observe any 
characteristic 1D Luttinger liquid signature in the collective charge 
density excitation spectral in one dimension; (9) the AP mode being 
strongly (and qualitatively) affected by the local field corrections 
(with the OP mode being relatively unaffected), in principle, it should 
be possible to investigate many body effects in 1D systems by studying 
the dispersion and damping of the AP mode (the OP mode on the other 
hand should be relatively insensitive to these many body corrections 
except at rather large wave vectors).

In our calculations we have concentrated on two distinct types of 
experimentally realizable two-component quasi-1D quantum plasmas 
confined in semiconductor (GaAs) quantum wire nanostructures: spatially 
separated doped quantum double-wire structures and photoexcited 
homogeneous 1D EHP in a single wire. Both systems are potentially 
interesting from the perspective of 1D collective excitation spectra 
and provide somewhat complementary information. Our hope is that our 
detailed quantitative (both numerical and analytical) investigation 
of the collective mode dispersion and spectral weight in two-component 
1D system will motivate experimental work (involving inelastic 
light scattering and far infrared frequency domain spectroscopies) on 
the subject searching for the 1D AP mode in quantum wire structures. 
 
\section*{ACKNOWLEDGMENTS}

This work is supported by U.S.-ONR and the U.S.-ARO.


\begin{figure}
\caption{ 
RPA plasmon dispersions of the 1D two-component system 
are shown for various impurity
scattering rates $\gamma$ as given in the figure. Here $\omega >0$
gives the dispersions of the mode and 
$\omega <  0$ the dampings of the mode. Thick (thin) lines
correspond to the OP (AP) mode. The overdampings of the
plasmons for $q<q_{c\pm}$ are clearly seen. 
We use the parameters: $\alpha = 1$ ($n_1 = n_2 =0.6\times
10^{6} \; {\rm cm}^{-1}$ and $m_1 = m_2 = 0.067m_e$)
$a=a_B$, and $d = 0.5 a_B$.
}
\label{fig2}
\end{figure}

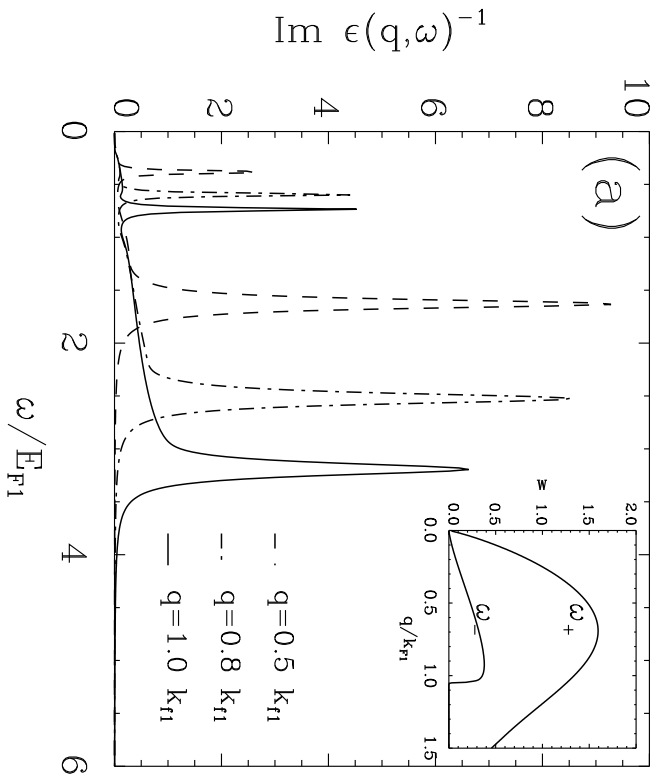
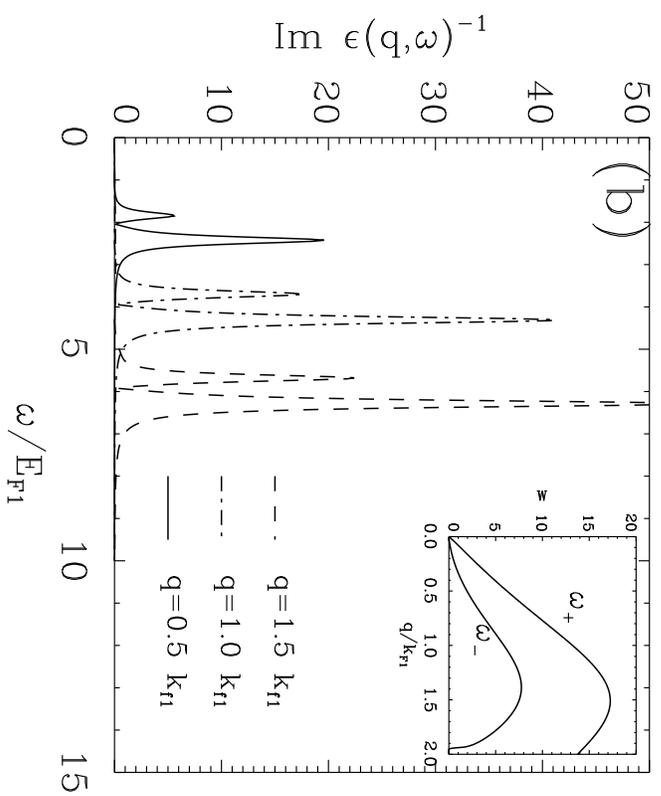
\begin{figure}
\caption{The RPA dynamical structure factor
Im$[\epsilon(q,\omega)^{-1}]$ in 1D two-component systems as a
function of frequency $\omega$ for different wave vectors $q$.
We use the parameters: (a) $n_1= 10^6
cm^{-1}$, $a=2 a_B$, $\alpha=0.2$, and $d=0$,  (b) for $n_1 = 0.6 \times 
10^6 cm^{-1}$, $a = a_B$, $\alpha = 0.9$, and $d=a_B$.  
We use the impurity scattering rate
$\gamma_1 = 0.1 E_{F1}$ and $\gamma_2 = 0.1 E_{F2}$ in both figures.
The inset shows the weights $W(q)$ of the plasmon modes in the absence
of the impurity scattering.
}
\label{fig3}
\end{figure}

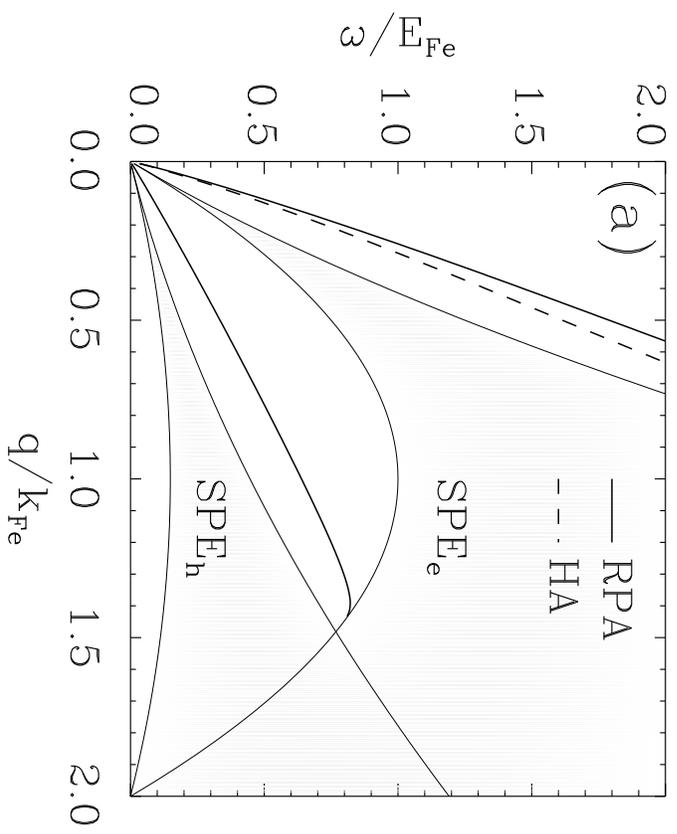
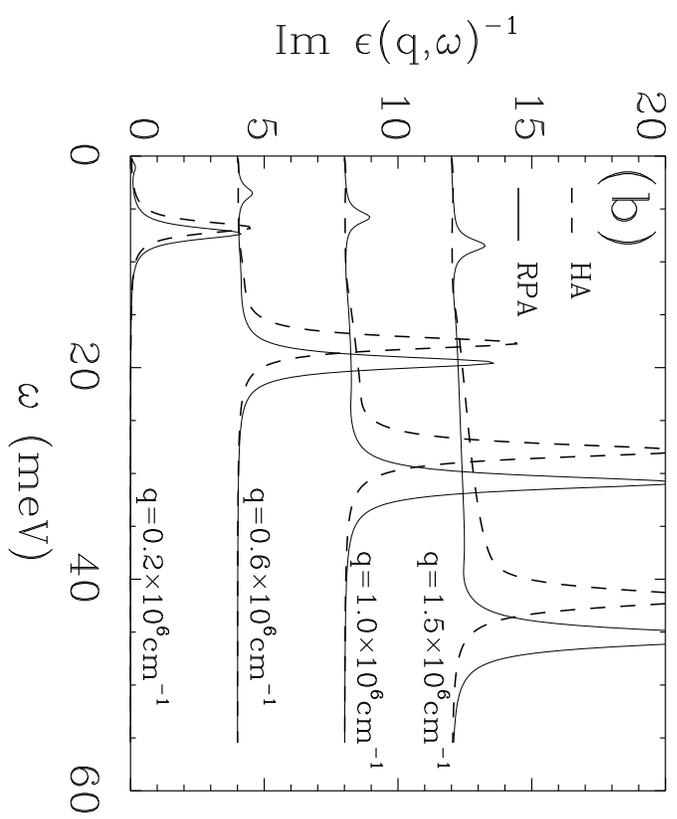
\begin{figure}
\caption{ The calculated
(a) plasmon dispersions and (b) spectral weights for densities $n_e = n_h =
1.0 \times 10^{6} \; {\rm cm}^{-1}$ in the photoexcited GaAs
quantum wire EHP within both RPA (solid lines) and HA (dashed
lines). The SPE$_{e,h}$ denote the single particle excitation of
the electron and the hole respectively. At this density the
AP mode $\omega_-$ is completely suppressed by the local
field effects in the HA (no dashed line corresponding to the AP mode 
exists within HA).
}
\label{fig4}
\end{figure}

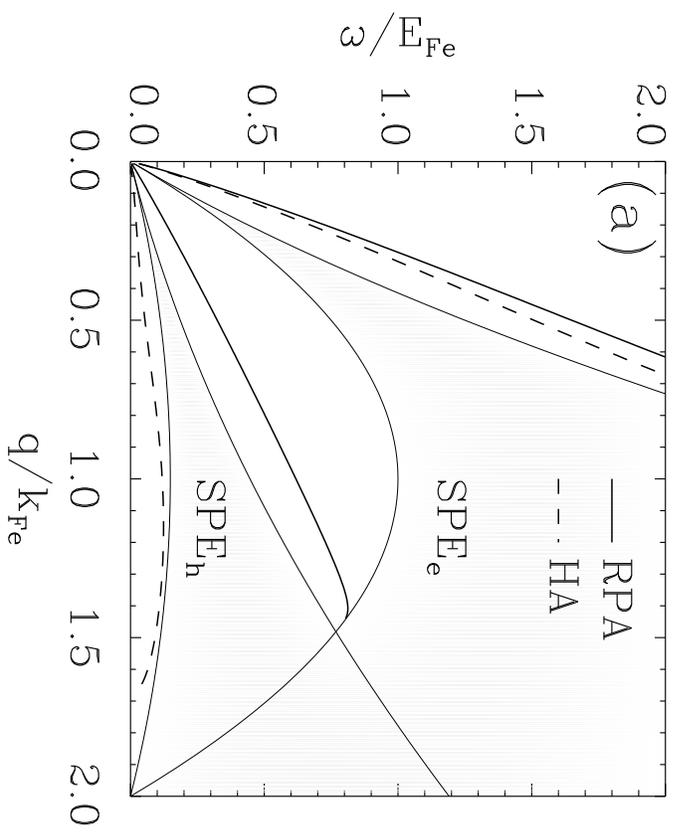
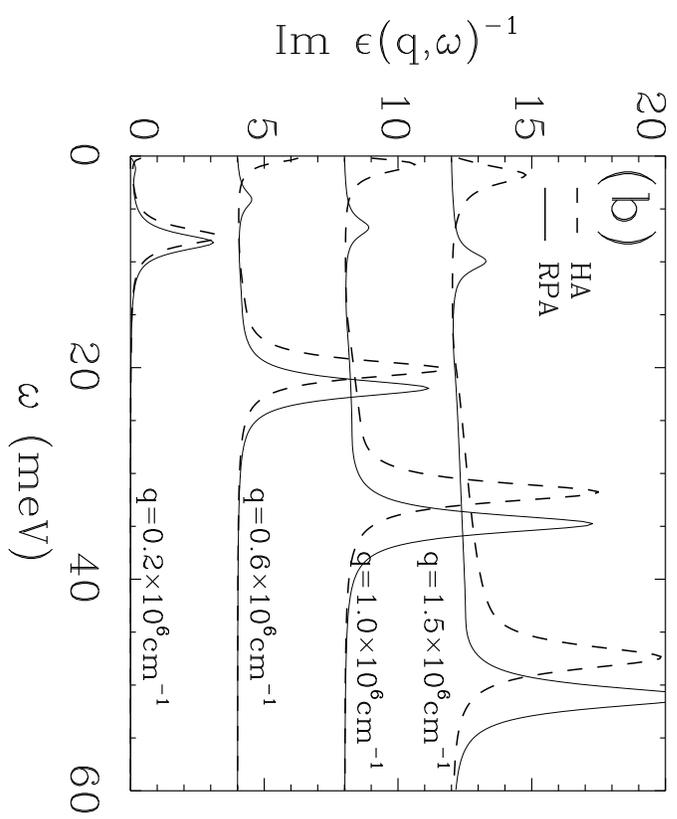
\begin{figure}
\caption{The same as Fig. \ref{fig4} for densities $n_e = n_h = 1.2
\times 10^{6} \; {\rm cm}^{-1}$. Note that the AP mode
$\omega_-$ lies below the SPE$_h$ due to the local field effects.
}
\label{fig5}
\end{figure}

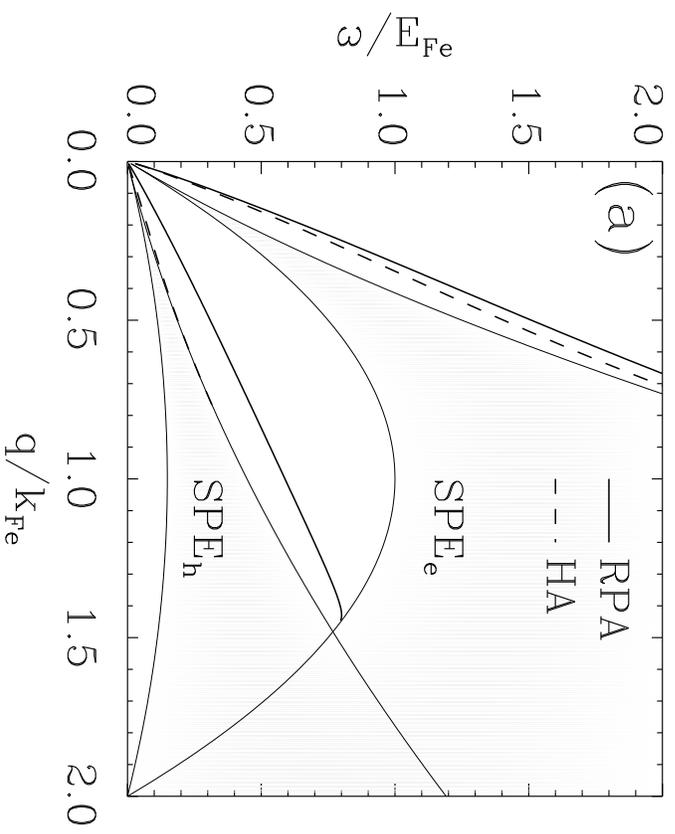
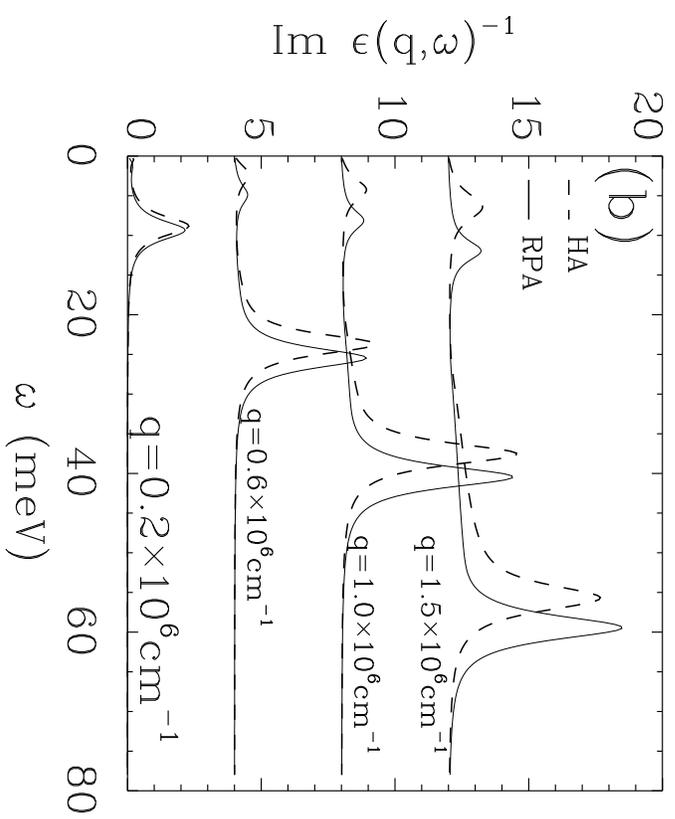
\begin{figure}
\caption{The same as Fig. \ref{fig4} for densities $n_e = n_h = 1.5
\times 10^{6} \; 
{\rm cm}^{-1}$. Note that the AP mode
$\omega_-$ lies between SPE$_e$  and SPE$_h$.
}
\label{fig6}
\end{figure}

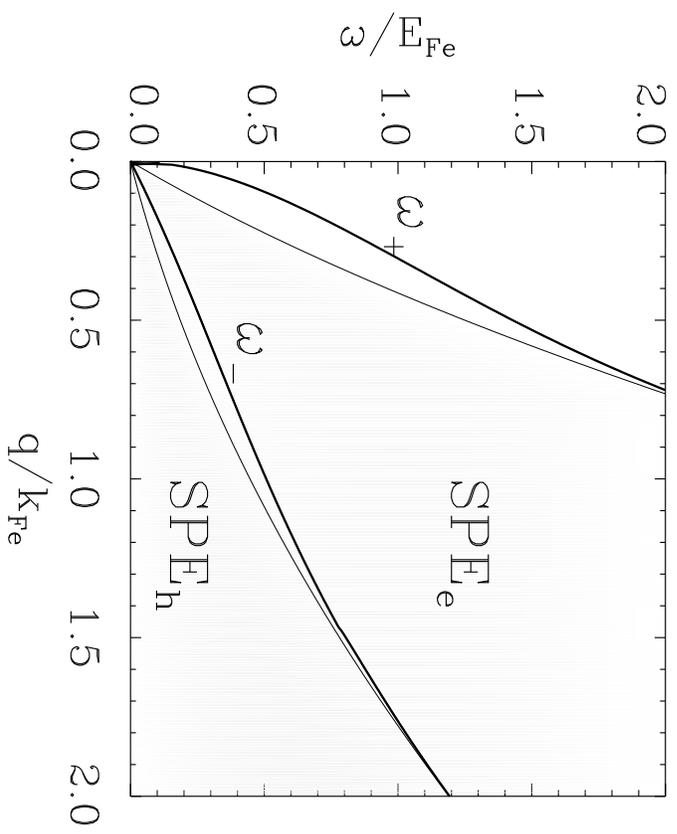
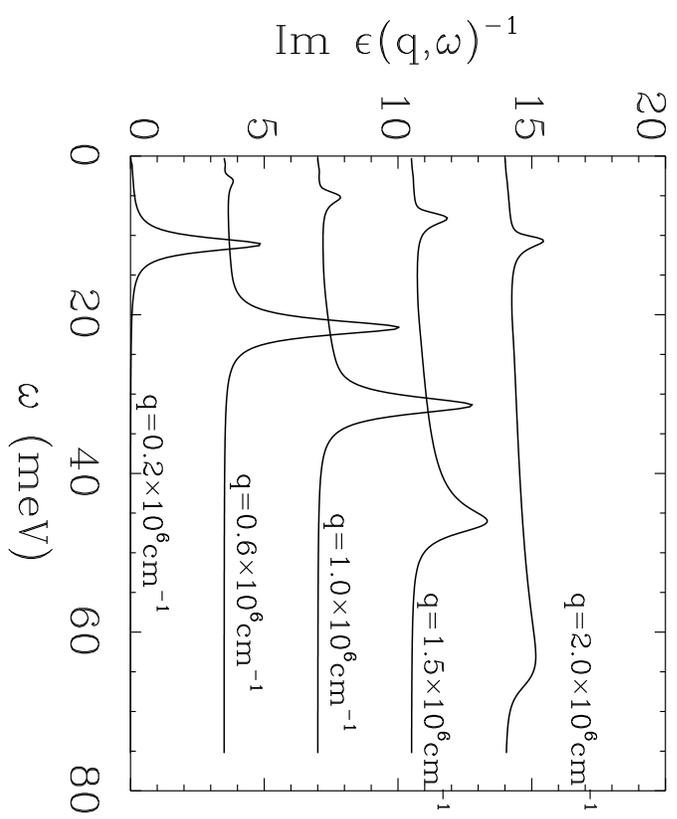
\begin{figure}
\caption{ The calculated
(a) plasmon dispersions and (b) spectral weights for densities $n_e = n_h =
0.6 \times 10^{12} \; {\rm cm}^{-2}$ in the photoexcited GaAs
2D EHP within RPA.
}
\label{fig8}
\end{figure}

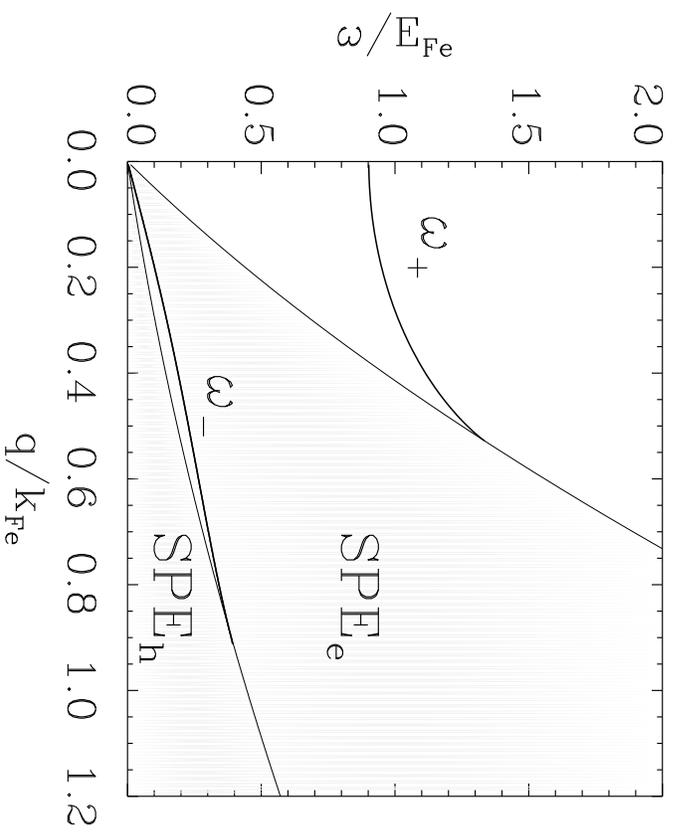
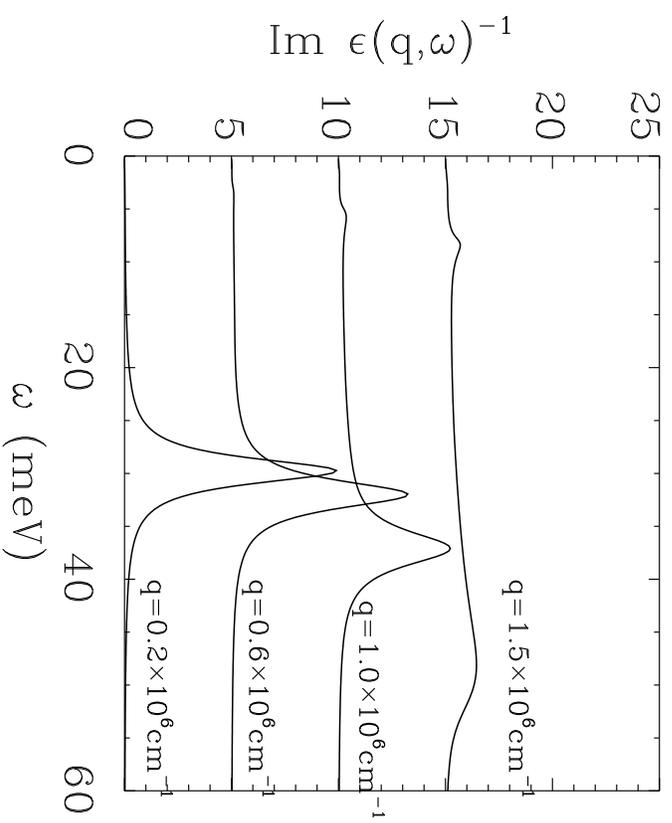
\begin{figure}
\caption{ The calculated
(a) plasmon dispersions and (b) spectral weights for densities $n_e = n_h =
0.465 \times 10^{18} \; {\rm cm}^{-3}$ in the photoexcited bulk GaAs
EHP within RPA.
}
\label{fig9}
\end{figure}


\begin{thebibliography}{99}

\bibitem{reed}For reviews see, e.g. 
{\it Nanostructures and Mesoscopic Systems}, edited by W. P.
Kirk and M. A. Reed (Academic, New York, 1992); C. W. J. Beenakker and
H. van Houten, in {\it Solid State Physics: Advances in Research and
Applications}, edited by H. Ehrenreich and D. Turnbull (Academic, New
York, 1991).

\bibitem{deme}T. Demel, D. Heitmann, P. Grambow, and K. Ploog,
Phys. Rev. B {\bf 38}, 12732 (1988); Phys. Rev. Lett. {\bf 66}, 2657
(1991); G. Hertel, {\it et al.}, Solid State Electron. {\bf 37}, 1289 (1994).

\bibitem{hans1}W. Hansen, J. P. Kotthaus, and U. Merkt, in {\it
Semiconductors and Semimetals}, Vol. 35, edited by R. K. Willardson
and A. C. Beer (Academic Press, San Diego, 1992).

\bibitem{pinc}A. Pinczuk and G. Abstreiter, in {\it Light Scattering
in Solids V}, edited by M. Cardona and G. G\"{u}ntherodt (Springer,
Berlin, 1989).

\bibitem{egel}T. Egeler, {\it et. al.}, Phys. Rev. Lett. {\bf 65}, 1804
(1990); R. Strenz, {\it et. al.}, {\it ibid}. {\bf 73},
3022 (1994); A. Schmeller, {\it et. al.}, Phys. Rev. B {\bf 49},
14778 (1994).


\bibitem{goni} A. R. Go\~{n}i, A. Pinczuk, J. S. Weiner, J. M.
Calleja, B. S. Dennis, L. N. Pfeiffer, and K. W. West, Phys. 
Rev. Lett. {\bf 67} 3298 (1991); A. R. Go\~{n}i, {\it et al.},
in {\it Phonons in Semiconductor Nanostructures},
p. 287, edited by J. P. Leburton, J. Pascual, and C. S. Torres
(Plenum, New York, 1993), p. 287. 


\bibitem{li} Q. P. Li and S. Das Sarma, Phys. Rev. B {\bf 43}, 11768
(1991); references therein.

\bibitem{li2} Q. P. Li, S. Das Sarma, and R. Joynt, Phys. Rev. B {\bf
45}, 13713 (1992).

\bibitem{dl}I. E. Dzyaloshinskii and A. I. Larkin, Zh. Eksp. Teor. Fiz.
{\bf 65}, 411 (1973) [Sov. Phys. JETP {\bf 38}, 202 (1974)].

\bibitem{pines} D. Pines, Can. J. Phys. {\bf 34}, 1379 (1956).

\bibitem{pinczuk} A. Pinczuk, J. Shah, and P. A. Wolff,
Phys. Rev. Lett. {\bf 47}, 
1487 (1981); references therein.

\bibitem{dsm} S. Das Sarma and A. Madhukar, Phys. Rev. B {\bf 23}, 805
(1981); J. K. Jain and S. Das Sarma, Phys. Rev. B {\bf 36}, 5949 (1987).

\bibitem{tan}B. Tanatar, Solid State Commun. {\bf 92}, 699 (1994).

\bibitem{hwang} E. H. Hwang and S. Das Sarma, Phys. Rev. B {\bf 50},
17267 (1994).

\bibitem{dashwang}S. Das Sarma and E. H. Hwang, Phys. Rev. B {\bf 54},
1936 (1996).

\bibitem{hu}Ben Yu-Kuang Hu and S. Das Sarma, Phys. Rev. Lett. {\bf
68}, 1750 (1992); Phys. Rev. B {\bf 48}, 5469 (1993).


\bibitem{liu}D. Z. Liu and S. Das Sarma, Phys. Rev. B {\bf 51}, 13821
(1995). 

\bibitem{mermin}N. D. Mermin, Phys. Rev. B {\bf 1}, 2362 (1970).

\bibitem{mahan}G. D. Mahan, {\it Many Particle Physics}, 2nd
ed. (Plenum, New York, 1990).


\bibitem{dassarma1} S. Das Sarma and E. H. Hwang, 
Phys. Rev. Lett. {\bf 81}, 4216 (1998).

\bibitem{weg}W. Wegscheider {\it et al}., Phys. Rev. Lett. {\bf 71},
4071 (1993).

\bibitem{pines1} D. Pines
and J. R. Schrieffer, Phys. Rev. {\bf 124}, 1387 (1961);
P. M. Platzmann, Phys. Rev. {\bf 139}, A379 (1965). 

\bibitem{platzmann} P. M. Platzmann and P. A. Wolff, {\it Waves and
Interactions in Solid State Plasma} (Academic, New York, 1973),
Chap. 5; M. V. Klein, in {\it Light Scattering in Solids I}, edited by
M. V. Cardona (Springer, Berlin, 1975), p. 147; J. Ruvalds,
Adv. Phys. {\bf 30}, 677 (1981).

\bibitem{alexeff} I. Alexeff and R. V. Neidigh, Phys. Rev. {\bf 129}, 516
(1961).


\bibitem{hal}F. D. M. Haldane, J. Phys. {\bf C14}, 2585 (1981).


\bibitem{sam} K. V. Samokhin, J. Phys.: Condens. Matter {\bf 10}, L533 (1998).

\end{thebibliography}
\end{document}